\documentclass{aa}  
\usepackage{ulem}
\usepackage{graphicx}
\usepackage{txfonts}
\usepackage{lipsum}
\usepackage{subcaption}         
\usepackage{lscape}
\usepackage{placeins}
\usepackage[colorlinks=true, allcolors=blue]{hyperref}
\usepackage{multirow}
\graphicspath{{Img_HD_142527/}}

\begin{document}

   \title{Exploration of the inner region of the system HD~142527\thanks{Based on observations collected at the European Southern Observatory under ESO programmes 095.C-0298, 096.C-0241, 097.C-0865 and 096.C-0248, 115.28BA.001 and 115.28BA.002.}}
    \titlerunning{Exploration of the inner region of the system HD~142527}

   \author{T. M. H. Tran\inst{1}\fnmsep\thanks{Email: \url{thi-my-hanh.tran@univ-lyon1.fr}}, M. Langlois\inst{1}, O. Flasseur\inst{1}, J.-C. Augereau\inst{2}, A. Boccaletti\inst{3}, P. Delorme\inst{2}, R. Gratton\inst{4}, A.~Kaszczyc\inst{1} and A.-M. Lagrange\inst{2,3}
        }
    \authorrunning{T. M. H. Tran et al.}

   \institute{Universite Claude Bernard Lyon 1, CRAL UMR5574, ENS de Lyon, CNRS, 69622 Villeurbanne, France
             \and Univ. Grenoble Alpes, CNRS, IPAG, F-38000 Grenoble, France
             \and LIRA, Observatoire de Paris, Université PSL, CNRS, Sorbonne Université, Université Paris Diderot, Sorbonne Paris Cité, CY Cergy Paris Université, 5 place Jules Janssen, 92195 Meudon, France
             \and INAF – Osservatorio Astronomico di Padova, Vicolo  dell’Osservatorio 5, 35122 Padova, Italy\\
            }
    
   \date{Received 30 January 2026 / Accepted 24 June 2026}
   \abstract
   {HD~142527 is a well-studied intermediate-mass T Tauri star surrounded by a transitional disk with a large dust cavity, spiral structures, and an accreting low-mass companion. Despite extensive observations, the system’s inner regions remain poorly understood, particularly regarding their influence on disk morphology and planet formation.}
   {This study aims to investigate the inner region of HD~142527 (separation < 50 au) with high detection sensitivity thanks to dedicated post-processing methods to search for undetected components and explore their potential role affecting the disk’s structure and evolution.} 
   {We analyze high-contrast imaging data obtained with the SPHERE instrument at the Very Large Telescope (VLT). To enhance detection sensitivity, we applied the \texttt{PACO} and \texttt{REXPACO} algorithms, dedicated respectively to the detection of point-like sources and to the reconstruction of circumstellar disks with high reliability, while exploiting both angular and spectral variations.}
   {We revisit the known companion HD~142527 B and update its photometry, astrometry, and accretion rate estimates. Furthermore, we identify a new candidate companion (CC) at an angular separation of $\rho \sim0.09''$ ($\sim$14 au). Though it is shown as a point-like object, the possibility that CC is a disk feature remains. Otherwise, it could be a young gas-giant planet or a brown dwarf with a mass of 12--50 $M\textsubscript{Jup}$. Additionally, we report the discovery of a tightly wound H$\alpha$ spiral feature in the inner disk, reconstructed for the first time by high contrast imaging. The spiral implies varying accretion dynamically linked to the known companion B and possibly to CC, suggesting ongoing interactions that influence the disk’s structure.}
   {Our findings provide new insights into the complex interactions within the HD~142527 system, highlighting the role of multiple companions in driving disk asymmetries and facilitating planet formation. Future high-resolution observations and dynamical modeling will be essential to fully understand the system's architecture and evolution.}

               
  \maketitle
\nolinenumbers

\section{Introduction}
\label{sec:intro}

Planet formation is closely related to the composition and evolution of protoplanetary disk material. Transition disks, characterized by significant gaps or inner holes, are of particular interest for the study of processes involved in the evolution of planets and disks. Their gaps are often attributed to the presence of forming planets, which can clear paths in the disk by accumulating material or dynamically affecting dust and gas \citep[e.g.,][]{linTidalInteractionProtostellar1993, dodson-robinsonTRANSITIONALDISKSSIGNPOSTS2011}. Understanding the properties and dynamics of transition disks can shed light on mechanisms behind planet formation and how the interactions of newly formed planets with their surrounding disk material affects their orbital characteristics and growth.

While current imaging infrastructures enable probing of the outer part of transition disks, resolving the inner region remains challenging. These inner parts, typically located within a few astronomical units from the central star, are critically important for understanding planet formation processes \citep{gravitycollaborationGRAVITYYoungStellar2019, kluskaFamilyPortraitDisk2020}. Although these regions may be reached using interferometry, in many cases it is insufficient due to the contrasts required to resolve faint structures \citep{avenhausExploringDustHD2017}. Pushing both the spatial and spectral resolving powers are hence a goal of attempts to discover and characterize exoplanets and circumstellar disks.

Due to its active circumstellar environment, many observations of the young star-disk system HD~142527 have been attempted throughout the past decades. Located at a distance of 159 $\pm$ 7 pc in the stellar association Scorpius-Centaurus (Sco-Cen; \citealt{gaiacollaborationGaiaDataRelease2023}), HD~142527 A is an accreting F6 III intermediate-mass T Tauri star with an age of 5.0 $\pm$ 1.5 Myr and a mass of 2.0 $\pm$ 0.3 \(M_\odot\) \citep{waelkensSWSObservationsYoung1996, mendigutiaSTELLARPARAMETERSACCRETION2014}. These stellar parameters place the system at a critical stage in planetary evolution where disk dispersal has taken place, marking the transition from active formation processes to mature planetary systems. We summarize the stellar properties of HD~142527 in Table \ref{tab:star}.

The star is surrounded by a highly asymmetric circumstellar disk located at a separation of 130--200 au \citep{fukagawaNearInfraredImagesProtoplanetary2006, rameauHighcontrastImagingClose2012, canovasNearinfraredImagingPolarimetry2013, christiaensSPIRALARMSDISK2014}. Several observations of this massive, optically thick, and almost face-on disk \citep{perezCOGASPROTOPLANETARY2014} have revealed complex features, including two noticeable shadows toward the north (position angle $\theta$ $\sim$ 0°) and southeast ($\theta$ $\sim$ 160°) in scattered light, and a number of spiral arms have been detected in both near-infrared (NIR) and submillimeter observations, indicating possible interactions with the inner region \citep{fukagawaNearInfraredImagesProtoplanetary2006, canovasNearinfraredImagingPolarimetry2013, christiaensSPIRALARMSDISK2014}.
Different estimates of the outer disk's position angle converge to $\theta_\text{out} \sim 160^\circ$, whereas its inclination has been reported with various values, from $i_\text{out} \approx$ 20--28$^\circ$ to 38$^\circ$ \citep{czekalaDegreeAlignmentCircumbinary2019, bohnProbingInnerOuter2022,nowakOrbitHD1425272024}.
The misaligned inner disk is assumed to lie between 0.3 and extend up to 30 au, and it is either geometrically flat and embedded in an optically thin halo \citep{verhoeffComplexCircumstellarEnvironment2011, avenhausExploringDustHD2017} or warped \citep{rosenfeldFASTRADIALFLOWS2014,casassusAccretionKinematicsWarped2015}, with $i_\text{in} \approx$ 20--33$^\circ$ and $\theta_\text{in}$ ranging from $-8^\circ$ to $20^\circ$ \citep{marinoSHADOWSCASTWARP2015, lazareffStructureHerbigAeBe2017, gravitycollaborationGRAVITYYoungStellar2019, scheuckAsymmetricStructureInner2026}. 
Between the two disks, the remarkably large dust cavity that extends from 30 to 130 au also gives a hint of ongoing planet formation processes \citep{casassusDYNAMICALLYDISRUPTEDGAP2012, marinoSHADOWSCASTWARP2015}. In addition, there is a substantial misalignment between the disks \citep{marinoSHADOWSCASTWARP2015, bohnProbingInnerOuter2022} that suggests a dynamical complexity in the surrounding environment. This has been confirmed by the fact that the system is actively accreting, with an estimated accretion rate of  $10^{-7} M_\odot \text{ yr}^{-1}$, which experiences significant variability on a timescale of 2--5 years \citep{lopezAccretionRatesHerbig2006, mendigutiaSTELLARPARAMETERSACCRETION2014}. This high accretion rate necessitates the existence of effective mass transport mechanisms across the immense disk gap \citep{casassusFlowsGasProtoplanetary2013}.

The discovery of HD~142527 B, a low-mass M-type companion with a mass of 0.13 ± 0.03 \(M_\odot\), located at a separation of $\sim$9--15 au, has provided significant insights into the disk's asymmetric morphology \citep{billerLikelyCloseinLowmass2012, closeDiscoveryHaEmission2014}. Observations suggest that this substellar companion plays a crucial role in shaping the disk's large-scale structures, including the formation of spiral arms and the clearing of the inner cavity \citep{fukagawaNearInfraredImagesProtoplanetary2006, priceCircumbinaryNotTransitional2018}. Despite extensive observational campaigns, several critical questions remain unanswered. Notably, hydrodynamical simulations suggest that HD~142527 B alone cannot account for the entire complexity of the disk's features \citep{liChallengeDirectImaging2024, nowakOrbitHD1425272024}.

One of the most intriguing aspects of the HD~142527 system is the presence of significant intensity nulls in the scattered light of the outer disk, which is attributed to an optically thick inner disk \citep{verhoeffComplexCircumstellarEnvironment2011, marinoSHADOWSCASTWARP2015}. To be able to cast these shadows, the inner disk is thought to be tilted by $\sim$70$^\circ$ compared to the outer disk \citep[inclination $i\sim$160\textdegree, $\theta$ $\sim \,$--20\textdegree]{casassusAccretionKinematicsWarped2015}.  Mid-infrared imaging and spectral energy distribution (SED) modeling suggest that the inner disk extends from 0.3 to 30 au \citep{verhoeffComplexCircumstellarEnvironment2011}. However, direct detection in scattered light has proven challenging. Early polarimetric differential imaging with the Nasmyth Adaptive Optics System Near-Infrared Imager and Spectrograph (NACO) at the VLT failed to detect the inner disk down to 0.10$''$ ($\sim$15 au; \citealt{avenhausStructuresProtoplanetaryDisk2014}). In the same year, \citet{rodigasPOLARIZEDLIGHTIMAGING2014} identified a highly polarized point source radially extended from B. At a separation of $\sim$17 au, however, it is not clear whether that detection is a multi-scattering artifact or a localized density enhancement of the inner disk signal. Near-infrared interferometric observations later found the inner disk to be consistent with a ring model of half-light radius of $\sim$0.2 au \citep{lazareffStructureHerbigAeBe2017, gravitycollaborationGRAVITYYoungStellar2019}. However, optical observations with SPHERE/ZIMPOL revealed an elongated dust structure at $\sim$0.15$''$ ($\sim$24 au; \citealt{avenhausExploringDustHD2017}), and the morphology of the structure fails to align with the shadows observed on the outer disk, raising questions about its nature and the extent of the inner disk.

Due to these unanswered questions, it cannot be ruled out that there are more complex interactions that have still not been observed, possibly due to unseen objects or multiple forming planets. Ongoing research aims to unravel these complexities and offers insights into how planetary systems evolve and the conditions under which they form.

In this article, we present new results on the inner region of HD~142527,  revealing a candidate companion (CC) and a number of extended features very close ($<$0.1$''$) to the host star. Section \ref{sec:obs} shows the data that we used and the data processing steps, while our results are presented in Sect. \ref{sec:results}.
The conclusions about this work are presented in Sect. \ref{sec:conclu}.

\begin{table}
\centering
\caption{\label{tab:star}Properties of HD~142527 A.}
\begin{tabular}{lcc}
\hline \hline
    Stellar properties & Value & Ref. \\ \hline
    \multirow{2}{*}{ICRS coord. } & $\alpha$: 15h 56m 41s.89 & \multirow{2}{*}{(1)} \\
                                            & $\delta$: \textminus42°19{${'}$}23{${''.}$}3 & {}\\ 
    $d$ (pc) & 159.3 $\pm$ 7.2 & (1) \\ 
    $\mu_\alpha \cos \delta$ (mas/yr) & \textminus10.924 $\pm$ 0.028 & (1) \\
    $\mu_\delta$ (mas/yr) & 26.149 $\pm$ 0.022 & (1)\\ 
    Spectral type & F6 III & (2) \\  
    Age (Myr) & 5.0 $\pm$ 1.5 & (3) \\  
    T\textsubscript{eff} (K) & 6500 $\pm$ 100, 6550 $\pm$ 100 & (3), (4) \\  
    $M_*$ (\(M_\odot\)) & 2.0 $\pm$ 0.3 & (3) \\  
    $L_*$ (\(L_\odot\)) & 16.3 $\pm$ 4.5 & (3) \\  
    $R_*$ (\(R_\odot\)) & 3.2 $\pm$ 0.2 & (4) \\  
    $A_V$ (mag) & 0.60 $\pm$ 0.05, 0.68 $\pm$ 0.04 & (5), (6) \\  
    $E(B-V)$ (mag) & 0.25 $\pm$ 0.04 & (3) \\  
    $G$ (mag)   &       8.1145 $\pm$ 0.0028     &       (1)     \\
    $J$ (mag) & 6.503 $\pm$ 0.029       & (7)   \\
    $H$ (mag) & 5.715 $\pm$ 0.031       &       (7)\\
    $Ks$        (mag) & 4.980 $\pm$ 0.020       &               (7) \\
    \hline
\end{tabular}
\\
\tablebib{(1) \citet{gaiacollaborationGaiaDataRelease2023}; (2) \citet{houkMichiganCatalogueTwodimensional1978}; (3) \citet{mendigutiaSTELLARPARAMETERSACCRETION2014}; (4) \citet{christiaensCharacterizationLowmassCompanion2018}; (5) \citet{verhoeffComplexCircumstellarEnvironment2011}; (6) \citet{bohnProbingInnerOuter2022}; (7) \citet{skrutskieTwoMicronAll2006}.
}
\end{table}

\section{Observations and data reduction}
\label{sec:obs}

\subsection{Observations}
\label{subsec:obs}
This study utilizes archival observations that were obtained in 2016, along with a new set of observations from early 2025, using the high-contrast imager and spectrograph SPHERE \citep{beuzitSPHEREExoplanetImager2019} of the Very Large Telescope at the Paranal Observatory in Chile. We process data from the three SPHERE subsystems, including the Integral Field Spectrograph \citep[IFS;][]{claudiSPHEREIFSSpectro2008}, the InfraRed Dual Imaging Spectrograph \citep[IRDIS;][]{dohlenInfraredDualImaging2008} and the Zurich IMaging POLarimeter \citep[ZIMPOL;][]{schmidSPHEREZIMPOLHigh2018}. All the data are non-coronagraphic and were taken in pupil-stabilized mode to make use of the angular combined with spectral differential imaging technique \citep[ASDI;][]{racineSpeckleNoiseDetection1999, maroisAngularDifferentialImaging2006}. This method exploits field rotation and spectral diversity of the observations to disentangle faint planetary or disk signals from stellar speckle noise, significantly enhancing contrast at small angular separations.
IRDIS/IFS data are taken simultaneously using IRDIFS observing mode, in which IFS operates in low-spectral resolution $YJH$ bands (950 to 1650 nm) while IRDIS exploits $K1-K2$ filters ($\lambda_{\rm c}$ = 2110 nm and 2251 nm, respectively). In the case of ZIMPOL, there are also two sets of simultaneous observations recorded by two separate detectors equipped with two filters, one of which is continuum ($CntHa$; $\lambda_{\rm c}$ = 644.9 nm) and another is either broadband ($B\_Ha$; $\lambda_{\rm c}$ = 655.6 nm) or narrowband ($N\_Ha$; $\lambda_{\rm c}$ = 656.53 nm) H$\alpha$.

\begin{figure}[ht]
    \subfloat{\includegraphics[width=\hsize]{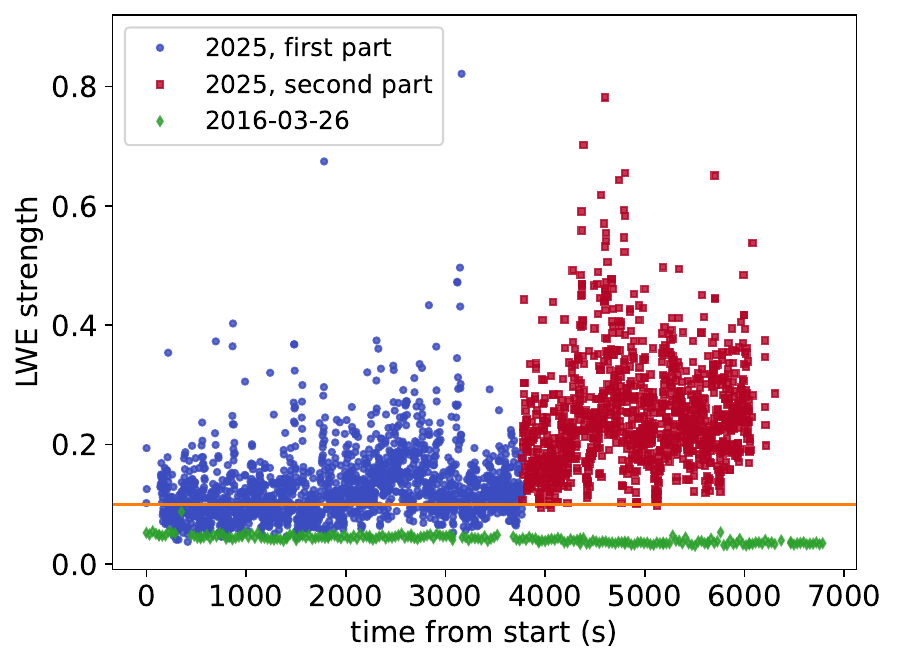}}
    \caption{\label{fig:LWE}Comparison of the LWE strength between the 2025 epoch and the best epoch on 26 March 2016 (green), calculated from differential tip-tilt sensor images. The 2025 epoch is divided into two parts: first 60\% (blue) and last 40\% (red) of the observing sequence. The orange horizontal line represents the minimum LWE strength value (10\%) to robustly claim the detection of LWE.}
\end{figure}

A summary of observations is presented in Appendix \ref{sec:app_obs}. 
In this list, IRDIS/IFS datasets were previously used by \citet{claudiSPHEREDynamicalSpectroscopic2019} while ZIMPOL data were mentioned by \citet{zurloWidestHaSurvey2020} and analyzed by \citet{cugnoSearchAccretingYoung2019}. Besides, a list of supplementary observations in data archive, unused in our analysis due to substantially lower sensitivity, can be found in Appendix \ref{sec:app_supplement_obs}.
As new features were discovered using archived data, we requested new SPHERE observations in similar observing setups in order to verify our findings. These observations were carried out in late April and early May 2025, under outstanding atmospheric conditions. However, IRDIS and IFS data were affected by the low wind effect (LWE), an artifact due to the imperfect adaptive optic correction, caused by refractive index variations when low altitude wind speed decreases \citep{sauvageLowWindEffect2015, sauvageTacklingLowWind2016, milliLowWindEffect2018}. In these data, the effect is substantially strong in the last 40\% of the observing sequence, degrading considerably the image quality. We calculated the LWE strength following \citet{milliLowWindEffect2018} using the differential tip-tilt sensor images corresponding to the observing sequence. This parameter provides a means to estimate the influence level of the LWE on the performance of the instrument. 
Our estimates of LWE strength for the 2025 epoch are illustrated in Fig. \ref{fig:LWE}. While the LWE strength for 60\% first frames of the observing sequence is 13.0 ± 6.2\%, indicating milder LWE, this value is approximately doubled in the last part, at 25.3 ± 9.2\%, with some frames severely damaged. Comparing to the 20–30\% Strehl loss driven by a 20\% LWE strength in SPHERE system estimated by  \citet{milliLowWindEffect2018}, these values show a significant deterioration of the instrument performance. For comparison, we also estimated the LWE strength of the best epoch on 26 March 2016, which yields only 4.2 ± 0.7\%, substantially under the conservative threshold of 10\% that asserts the LWE presence.

\subsection{Data processing}

To optimize the detection of faint companions and disk structures, we applied dedicated post-processing algorithms, \texttt{PACO} \citep{flasseurExoplanetDetectionAngular2018, flasseurRobustnessBadFrames2020} and \texttt{REXPACO} \citep{flasseurREXPACOAlgorithmHigh2021}, which are unsupervised learning techniques specified for angular and spectral differential imaging data. 
\texttt{PACO} is a state-of-the-art method that statistically models spatial correlations in background noise, allowing photometric preservation of point sources. \texttt{REXPACO} extends this framework by incorporating extended image reconstruction and deconvolution to reconstruct disk morphologies with high reliability.
Each of these algorithms has an extension, ASDI \citep{flasseurPACOASDIAlgorithm2020, flasseurREXPACOASDIJoint2024}, that combines angular and spectral differential imaging to take advantage of both differential imaging approaches, thus allowing one to achieve a higher signal-to-noise ratio (S/N), reach higher contrast limits, and reconstruct structures at smaller separations. \citet{chomezPreparationUnsupervisedMassive2023} studied a sample of 24 stellar systems during 200 nights of SPHERE archival data with \texttt{PACO} ASDI and found that the obtained contrast limits were significantly improved compared to more classical algorithms, notably at a level of ten times at separations between 0.2 and 0.5${''}$. This improved sensitivity was also confirmed in the SHINE survey of around 650 SPHERE datasets \citep{chomezSPHEREInfraredSurvey2025}.
In the work presented in this paper, we applied \texttt{PACO} and \texttt{REXPACO} to consistently process data from the three SPHERE/VLT instruments, IRDIS, IFS and ZIMPOL, after standard pre-processing pipelines.

The procedure of basic data pre-processing includes remapping, bad pixel removal, bias (for ZIMPOL only), background (for IRDIS and IFS), dark subtraction, and flat-field correction. ZIMPOL data were reduced using the Interactive Data Language (IDL) ZIMPOL software pipeline developed at ETH Zurich \citep{schmidSPHEREZIMPOLHigh2018}, while IFS data reduction was carried out by the SPHERE Data Center \citep{delormeSPHEREDataCenter2017}. In the case of IRDIS data, depending on availability, the reduction of some datasets was performed by the SPHERE Data Center, others with the \texttt{ScientificDetectors} package\footnote{\url{https://github.com/emmt/ScientificDetectors.jl}}. 
Additionally, these pipelines provide the parallactic angles that indicate the field angular rotation of the images with respect to the star. These field rotation angles were calculated using the information provided in headers. The images were then aligned and centered by fitting a two-dimensional Gaussian profile to the stellar point spread functions (PSFs), except special cases as detailed in Sect. \ref{subsec:saturation}. All data sets were also corrected for anamorphism to compensate for image distortions due to common path optics \citep{maireSPHEREIRDISIFS2016, schmidSPHEREZIMPOLHigh2018}. Especially for IFS observation of 2025 epoch, we manually selected frames to avoid several critically impaired frames due to LWE, as mentioned earlier in Sect. \ref{subsec:obs}. 
The step reduces the cube size by about 10\%, but has little effect on the field rotation.
This selection process is not required on IRDIS 2025 dataset because the LWE is less pronounced on longer wavelengths, and on the other hand, it is beneficial to maintain the number of frames to compromise between background statistics and the impacts introduced by LWE.

IRDIS and IFS data consist of two datasets recorded each night, which are different from each other in total integration time and detector integration time (DIT) and the use of neutral density (ND) filters. 
The highest S/N is obtained by processing the saturated data because the use of stronger ND filters during long integration times reduces the signal level making the relative contribution of noise more significant including high thermal background noise level. Consequently, we did not employ the unsaturated datasets as the main cubes to implement \texttt{PACO}/\texttt{REXPACO}. Instead, the data cubes that were post-processed are the saturated ones. With a customized routine to address the saturation issue explained in Sect. \ref{subsec:saturation}, these data offer better background structure capture by \texttt{PACO} and \texttt{REXPACO}'s model. In addition, we still made use of the unsaturated datasets for centering and photometry calibration, which is explained in detail in the next section.

\begin{figure*}[ht]
    \centering
    \subfloat[]{\includegraphics[width=0.58\textwidth]{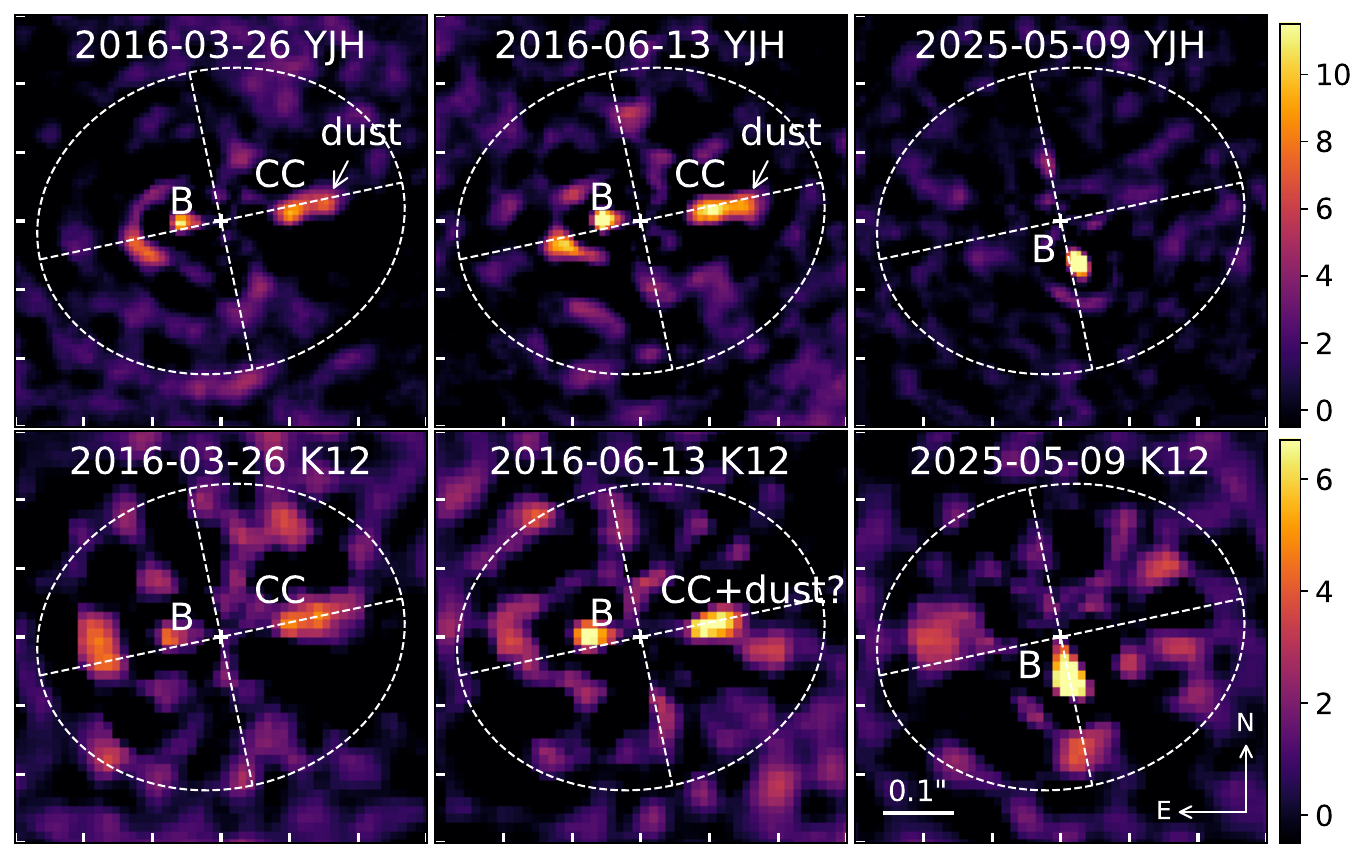}}
    \hfill
    \subfloat[]{\includegraphics[width=.408\textwidth]{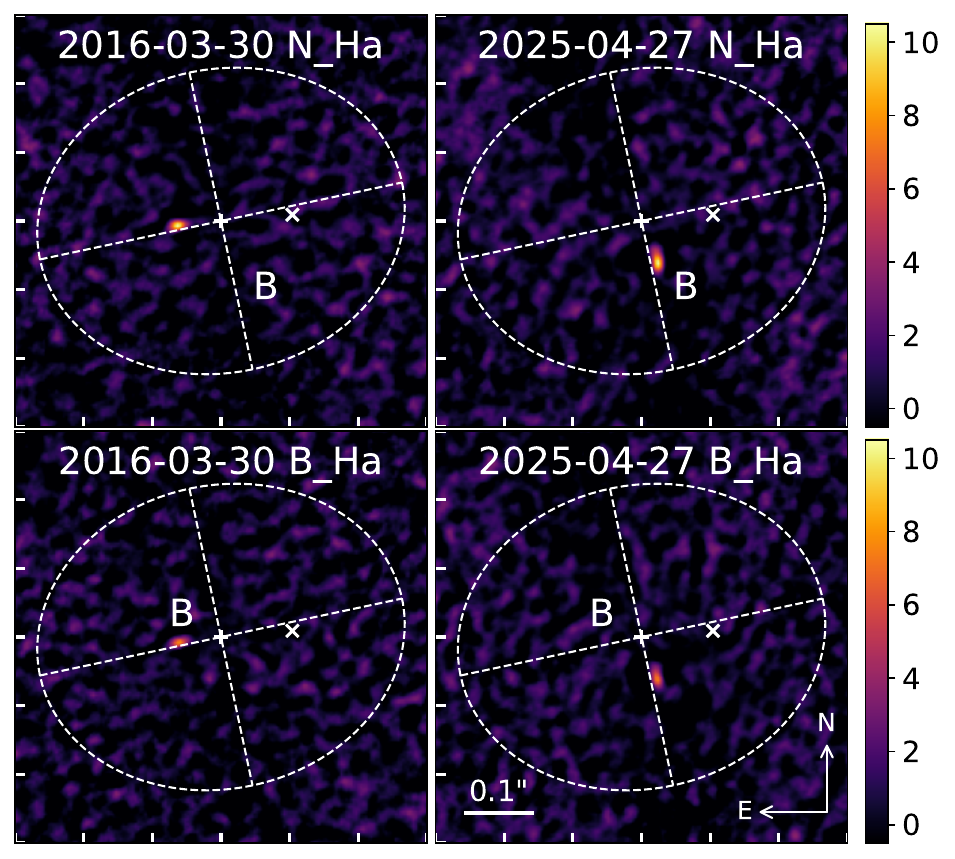}}
    \caption{\textit{Panel (a)}: \texttt{PACO} ASDI S/N detection maps of IRDIS and IFS observations combined with equal weight across all spectral bands, obtained by \texttt{PACO} ASDI. The dwarf companion HD~142527 B appears in all images with a significant S/N. In the two 2016 epochs, we identified the CC close to an extended signal, detected with a high S/N. In all the images, the star is marked with a white cross. An ellipse of the same spatial extent is overlaid in all panels for visual comparison purposes. \textit{Panel (b)}: Same as (a) but for ZIMPOL observations. The white $\times$ symbol marks the detected position of the CC on 26 March 2016, five days before the ZIMPOL observations.}
    \label{fig:snr_maps}
\end{figure*}

\subsection{Dealing with saturation}
\label{subsec:saturation}
In ASDI approaches, image sequences must be well centered in order to assess and isolate the noise contribution. Since data are all non-coronagraphic, our recentering relies on fitting a two-dimensional Gaussian profile to the PSF of the star in each image and shifting these images the common geometrical center. 
Nevertheless, this process is not suitable for saturated data taken in IRDIS $K12$ in which the PSF cores are deformed. In this case, a special centering procedure was applied. Even though the linearity is not guaranteed, the PSF peaks still indicate the position of the star within a tolerance less than 0.05 pixel. Therefore, we used a robust Gaussian fit \citep{huberRobustEstimationLocation1964} on the images of the $K2$ filter to estimate the star position, then shifted the images of both the cameras to this point with the offset between the two images taken into account by cross-correlation. In the case of IFS images, the saturation is very mild, the PSF cores still indicate the center of the rotation with uncertainties less than 0.05 pixel, and hence the centering was not affected.

Saturation also raises an issue for the photometric calibration. In the case of our IFS data, we used the unsaturated dataset of each night to represent the star's PSF of that night's observations. However, we did not apply the same procedure to IRDIS data since the quality of the unsaturated $K12$ sets is significantly impacted by thermal emission. Only a few pixels, whose values are saturated, are present at the center of the PSF images. We identified their locations to mask the saturated regions while fitting a modeled Airy pattern to the remaining unsaturated regions of the PSF. The masked region's values were substituted with those from the fit modelled data in the original dataset. The revised PSFs therefore have their peak values synthesized, while their wings remain unchanged.

\section{Results and discussion}
\label{sec:results}

\subsection{Detection and reconstruction}

\begin{figure*}[ht]
    \subfloat[]{\includegraphics[width=.595\textwidth]{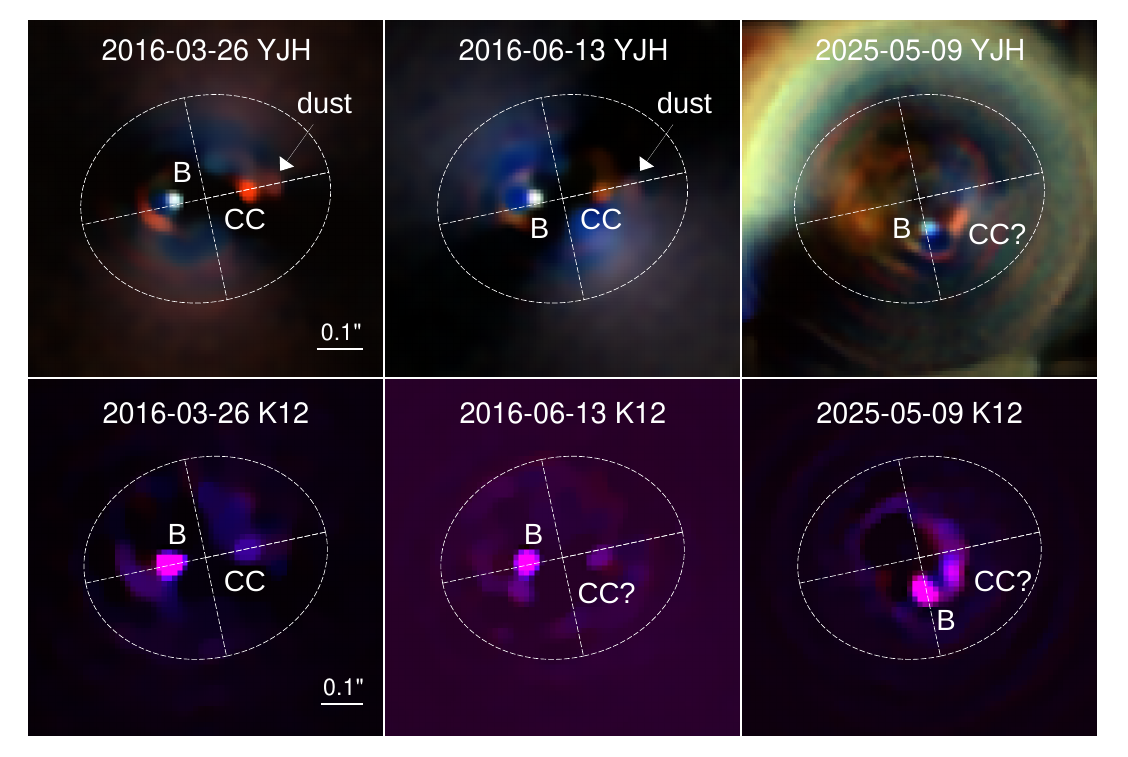}
        \label{fig:reconstruction_irdifs}
        }
    \hfill    
    \subfloat[]{\includegraphics[width=.405\textwidth]{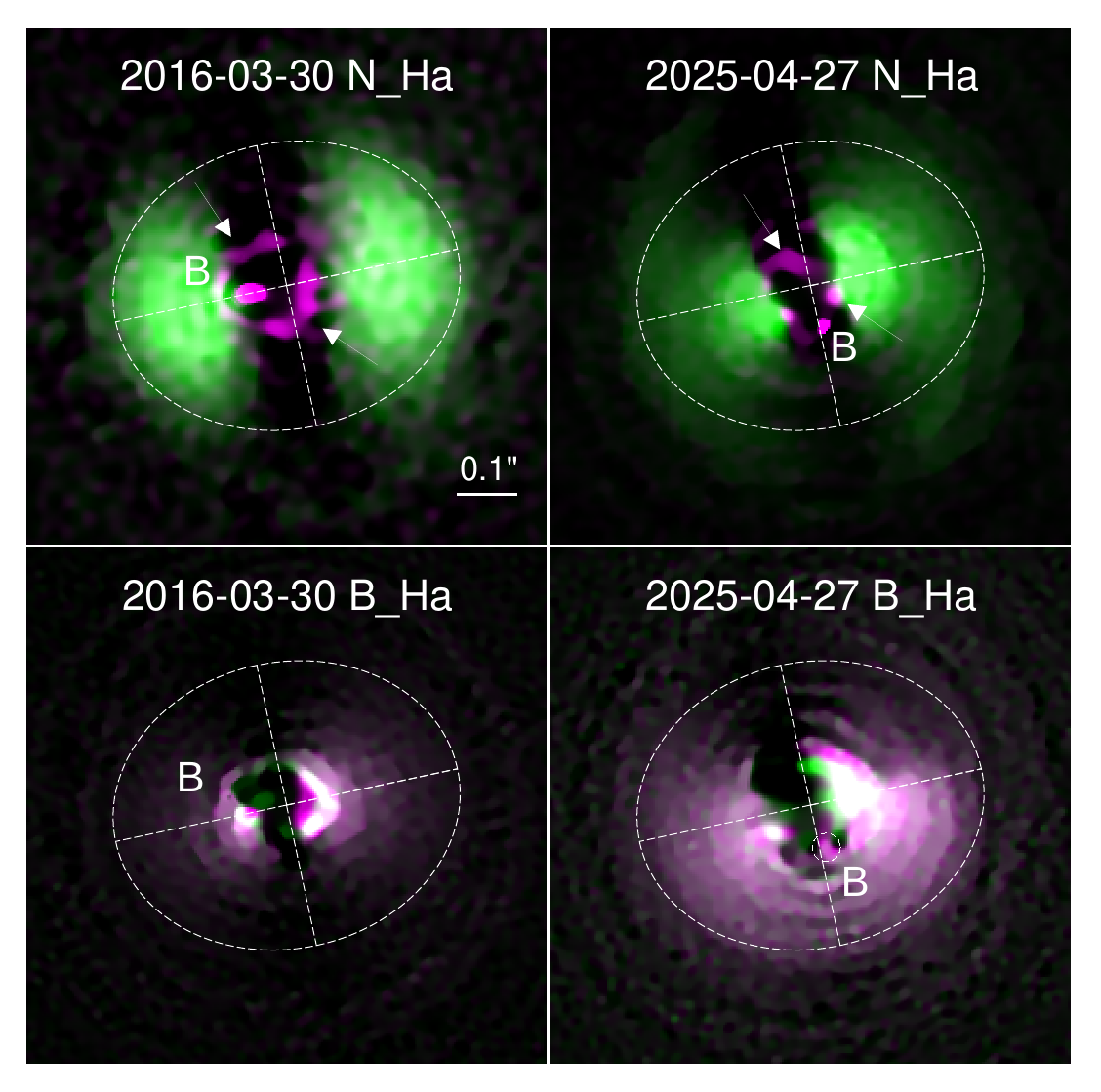}
        \label{fig:reconstruction_zp}
        }
    
    \caption{\textit{Panel (a)}: Enlarged deconvolved reconstructed images using \texttt{REXPACO} toward the central area by IFS and IRDIS instruments in linear scale; the units are arbitrary. Colors encode different wavelengths, from blue (shorter wavelengths) to red (longer wavelengths). In the IRDIS images, B is oversaturated to emphasize fainter structures.
    \textit{Panel (b)}: ZIMPOL's reconstructed flux distribution toward the central area, scaled to bandwidths of the H$\alpha$ line filters. Colors encode different wavelengths, showing continuum (green) and line (magenta) emission. Images are shown in linear scale. The units are arbitrary. B is oversaturated to emphasize fainter structures. Arrows point to spiral-like features that are discussed later in Sect. \ref{sec:spiral}.
    The CC is not seen in the H$\alpha$ filters, as argued in the text.
    In all the panels, an ellipse of identical spatial extent, centered at the star, is marked for visual comparison purposes. Intensity scales of images are only comparable across different epochs of the same observing filters. 
    }
    \label{fig:reconstruction}
\end{figure*}

\begin{figure*}[ht]
    \centering
    \subfloat{\includegraphics[width=0.525\textwidth]{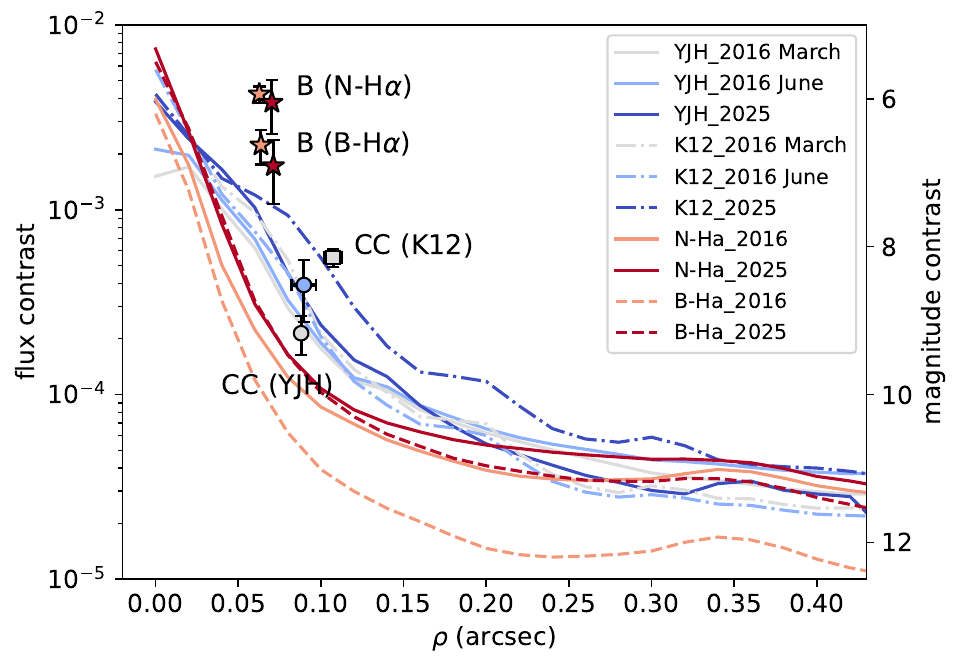}}
    \hfill
    \subfloat{\includegraphics[width=0.475\textwidth]{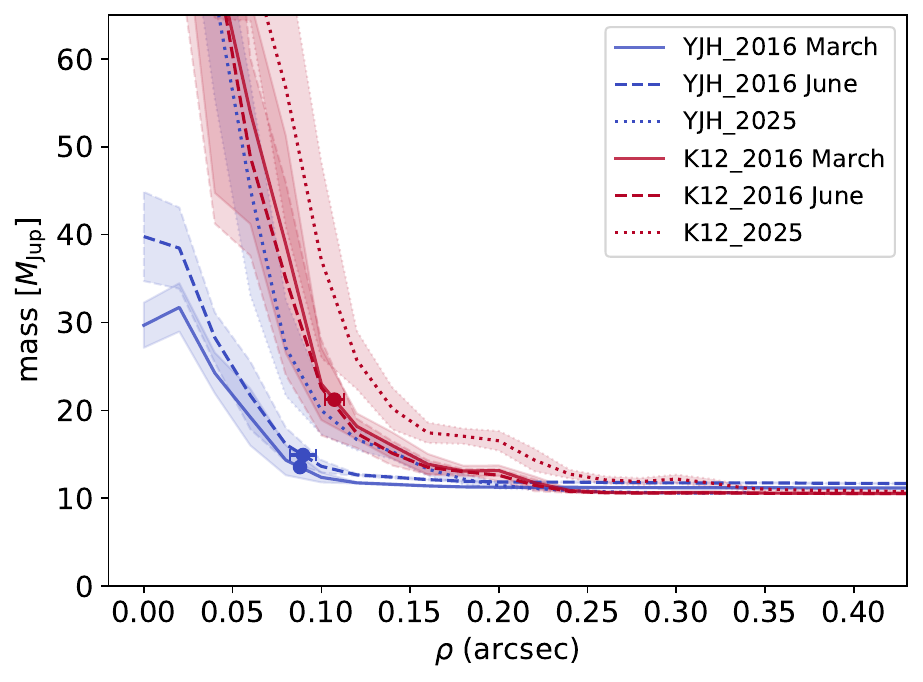}}
    \caption{\label{fig:contrast_curves}\textit{Left}: Five sigma contrast curves of all observations mentioned in this study overlaid with 3$\sigma$ contrast uncertainties of B and CC. \textit{Right}: Mass limits obtained from NIR observations, assuming a system age of 5.0 Myr. Colored dots in the two epochs in 2016 show the intersections between the detected separations of the CC with a significance level of 3$\sigma$ and the mass limit curves in the corresponding IFS or IRDIS filters.
    This information is not displayed for B because its mass falls far above the minimum in the IRDIS and IFS observations. The separation uncertainties, given with a 3$\sigma$ level, allows the consistency of the measurements in separation to be validated.}
\end{figure*}

Figure \ref{fig:snr_maps} shows \texttt{PACO} ASDI S/N maps of detection for all epochs in the inner region of the HD~142527 system obtained by different instruments: IRDIS, IFS and ZIMPOL. Values of S/N in each case can be found in Table \ref{tab:astrometry}. We confirm the detection of HD~142527 B with all instruments and for all observational epochs (S/N > 5 in almost all cases), with significant H$\alpha$ line emission (observed by ZIMPOL) indicating ongoing accretion. From 2016 to 2025, the low-mass companion moves in the counterclockwise direction from east to west passing by north. Correspondingly, we plotted the \texttt{REXPACO} reconstructed fluxes of the inner region of the system in Fig. \ref{fig:reconstruction}. Here B also appears distinctly in all the NIR observations and line H$\alpha$ panels, consistent to the detection maps.
As a side note, to facilitate the comparison among panels, we plotted an ellipse constructed by visually fitting the observed continuum H$\alpha$'s brightest region around the star that covers all the features of interest, then overlay it on all the images in Figs. \ref{fig:snr_maps} and \ref{fig:reconstruction}. Also, there appears a bright region outside the ellipse in the 2025 IFS reconstructed map, but it is just an artifact introduced by regularization at the edges of the chosen region. 

In addition, we discovered a signal in the NIR at $\sim$0.09${''}$ to the west of the host star during the 2016 epochs. The signal is absent in ZIMPOL H$\alpha$ filters during this period but detected with high S/N (4.8--11.7) in both IRDIS and IFS, showing a significant red color. We refer to this detection as a CC. If the signal actually indicates a companion to the star, the companion is likely embedded in opaque dust, making it challenging to detect at visible wavelengths. Nevertheless, with given information, it cannot be ruled out that it could be a disk feature.

In $YJH$ filters, B is detected at $\sim$0.06${''}$ with a mean contrast of 1.21$\times{10^{-2}}$ while CC is found at $\sim$0.09${''}$ and a contrast of 1.24$\times{10^{-4}}$, which is around 100 times fainter. We also notice a signal in the vicinity of CC that appears to be another point source in the detection maps. This signal is found in both IRDIS and IFS observations of both the 2016 epochs. However, it is likely material from a disk because it appears to evolve rapidly and significantly across spectral channels. Its spectrum also does not resemble a compact object (Sect. \ref{subsec:photometry}). Therefore, we marked this detection as ``dust'' in Fig. \ref{fig:snr_maps}. 

Attempting to look for CC again in a third epoch, we requested more observations of CC in 2025. However, CC is not unambiguously detected in the 2025 epoch. Instead, we retrieve a signal in the \texttt{REXPACO} reconstruction maps that aligns with the very red color of CC, though without a clear detection, we are unable to conclude whether it comes from the same CC or not. Besides, it is noticeable that the signal appear to be elongated, which possibly implies the CC-dust combination seen in 2016 epochs, or it could simply indicate a disk feature. We note, however, that in all the reconstructed images, it seems to stand out against the environment with its red color.
The LWE effect mentioned above certainly deteriorated the image quality of the 2025 epoch; nevertheless, the primary reason for the detection's difficulty might be due to the abnormal shape of the CC resulting from the actively dusty region, which is retrievable by \texttt{REXPACO} but not \texttt{PACO}, because the latter imposes a PSF-like shape for the detection.

Using \texttt{PACO}, we estimated the detection limits of the observations which are presented by 1-D contrast curves, as shown in Fig. \ref{fig:contrast_curves}. 
While ZIMPOL curves represent the mean sensitivity of both continuum and H$\alpha$ bands, IRDIS and IFS curves are combined across spectral channels with the weighting factors corresponding to the CC's measured contrast in the best epoch of March 2016.
Generally, it is noticeable that, at separations $\lesssim0.1{''}$, shorter-wavelength observations provide higher sensitivity, although substellar and planet sources tend to emit or reflect less flux in this wavelength domain than in NIR. In NIR, the best sensitivity was achieved during the 2016 epochs, where we estimate contrast limits down to $\sim$9 magnitudes at an angular separation of 
0.1${''}$.

Furthermore, it is discernible that, compared to the common approach using principal component analysis for PSF subtraction, at a separation of 0.1${''}$, \texttt{PACO} achieves contrast limits of $\sim$3--5 magnitudes higher in IFS/IRDIS filters and 1--2 magnitudes higher in ZIMPOL H$\alpha$ \citep{claudiSPHEREDynamicalSpectroscopic2019, cugnoSearchAccretingYoung2019}. As discussed by \citet{cugnoSearchAccretingYoung2019}, the contrast performance seems to be better in broadband H$\alpha$ due to more overall fluxes obtained in this band, while detection of H$\alpha$ line emission is preferred in narrowband H$\alpha$. Our new analysis achieves higher S/N in narrow H$\alpha$ band, as expected.

\begin{table*}[ht]
\caption{\label{tab:astrometry}Signal-to-noise ratios of the observations and astrometric measurements of the detected sources extracted from \texttt{PACO}.}
\centering
\begin{tabular}{lcccccccc}
\hline \hline

        \multirow{2}{*}{UT Date} & \multirow{2}{*}{Filters} & \multicolumn{3}{c}{B} & \multicolumn{3}{c}{CC} \\
        ~ & ~ & S/N & $\rho$ (mas) & $\theta$ (deg) & S/N & $\rho$ (mas) & $\theta$ (deg) \\ \hline
        
        \multirow{2}{*}{2016-03-26} & $YJH$ & 11.3 & 62.0 ±  0.2 & 95.58 ± 0.15 & 9.6 & 88.0 ± 0.5 & 281.00 ± 0.31 \\
        & $K12$ & 4.3 & 61.4 ± 0.7 & 95.15 ± 1.13 & 4.8 & 107.4 ± 1.8 & 273.32 ± 0.97 \\
        
        \multirow{2}{*}{2016-03-30} & narrow-H$\alpha$ & 10.6 & 63.7 ± 0.2 & 98.89 ± 0.61  & ... & ... & ... \\
        & broad-H$\alpha$ & 7.7 & 62.7 ± 0.2 & 100.96 ± 0.59  & ... & ... & --  \\
        
        \multirow{2}{*}{2016-06-13} & $YJH$ & 18.1 & 56.8 ± 0.3 & 92.70 ± 0.25 & 11.7 & 89.7 ± 2.5 & 275.22 ± 1.63 \\
        & $K12$ & 8.1 & 62.5 ± 0.5 & 91.53 ± 0.43& 8.1\tablefootmark{a} & ... & ... \\
        
        \multirow{2}{*}{2025-04-27} & narrow-H$\alpha$ & 10.5 & 70.3 ± 0.3 & 199.12 ± 0.61& ... & ... & ... \\
        & broad-H$\alpha$ & 7.5 & 71.2 ± 0.3 & 199.32 ± 0.63& ... & ... & ... \\
        
        \multirow{2}{*}{2025-05-09} & $YJH$ & 21.1 & 69.9 ± 0.3 & 194.89 ± 0.19& 3.8\tablefootmark{**}  & 91.1 ± 0.5\tablefootmark{***} & 231.84 ± 0.29\tablefootmark{***} \\
        & $K12$ & 13.4 & 81.8 ± 1.4 & 191.63 ± 0.96& 3.0 & 111.5 ± 1.8\tablefootmark{***} & 247.57 ± 0.98\tablefootmark{***} \\
    \hline
    \end{tabular}
    \tablefoot{All the values of astrometric parameters are accounted for error budget mentioned in the text.
    \tablefoottext{a} {Contaminated by dust signal.} 
    \tablefoottext{b} {Only in this case, S/N obtained by combining values of different spectral channels with weights that are the CC's normalized spectrum found in 2016 March. In other cases, the S/N values are equally combined across spectral channels.} 
    \tablefoottext{c} {Coordinates estimated for the red signal in the 2025 epoch, which still needs verification to confirm if it is from the CC. Measurements of CC, especially in the 2025 epoch, should be taken with caution because they have been retrieved by assuming the detection is point-like, while it is not the case due to the material around CC.}}
\end{table*}

We employed the \texttt{MADYS} package \citep{squicciariniMADYSManifoldAge2022} to convert the instruments' sensitivity into mass detection limits. In order to account for contrast variations in azimuthal direction, especially at small angular separations where speckle noise dominates, we used the two-dimensional contrast maps produced by \texttt{PACO} for both IRDIS and IFS. Thanks to BT-Settl isochrone model \citep{baraffeNewEvolutionaryModels2015}, we found mass limits corresponding to the contrast maps. The resulting mass limit curves are obtained by azimuthally averaging the mass limit maps in 0.02$^{\prime\prime}$-wide annuli. In the right panel of Fig. \ref{fig:contrast_curves}, the mass limits for different epochs are shown with the shaded regions representing the standard deviation within each annulus. One can see that, at the separations where CC (marked as separated dots for each epoch) is detected in IRDIS and IFS in two epochs in 2016, the minimum detectable mass is around 12--20 $M\textsubscript{Jup}$.

\subsection{Photometry and astrometry analysis}
\label{sec:photo-astro}
\subsubsection{Photometry}
\label{subsec:photometry}
Due to the complex and dusty environment, particularly the dust signal tightly bound to CC, we performed a customized \texttt{PACO} "unmixing" run in order to retrieve reliable astrometric and photometric measurements of CC and the dust signal in the 2016 epochs. This approach is similar to the one introduced by \citet{flasseurREXPACOASDIJoint2024}, which is devoted to distinguish between the contribution of point-like sources and that of extended features. First, we estimated the position and flux of CC and the dust component separately. These measurements are, however, biased because the small angular separation between the two components hinders accurate disentanglement. To address this, we iteratively suppressed the signal of one component while re-estimating the parameters of both. This two-step process was repeated multiple times until the measurements converged to stable values. By doing this, we disentangled the signals of the CC and dust, obtaining more robust estimates. Nevertheless, we acknowledge that this approach is ineffective to remove bias in the photometry i.e in contrast when dust lies along the line of sight. In the case of HD~142527, an inner disk is expected at these separations (see Sect. \ref{sec:intro}). Therefore, our measurements possibly represent only the upper limits, especially at longer wavelengths where the impact of thermal emission is more significant.

Photometric values are extracted for each spectral band in form of contrast as a ratio between fluxes of the off-axis objects and the central star. Considering the relevance of the wavelength, only values of spectral channels with S/N more than 1 are considered.
We adopt the scheme of evaluating photometric error budget presented by \citet[Sect.~4.3.2]{chomezPreparationUnsupervisedMassive2023}. However, since our observations are without coronagraphs, the star photometric uncertainties are directly retrieved from the PSFs and combined quadratically with the \texttt{PACO} photometric errors to obtain the total uncertainties. In addition to IRDIS $K1$ and $K2$ contrast values, those of $Y, J$ and $H$ bands from IFS are separately computed by
taking the median of spectral bands in three ranges: below 1.15 $\mu$m, from 1.15 to 1.35 $\mu$m, above 1.35 $\mu$m, respectively.

To convert the contrast measurements into physical flux units, first, we obtained a synthetic spectrum of the star HD~142527 A in 2MASS system by applying its filter transmissions and a spectrum of Vega \citep{bohlinHSTStellarStandards2007}, using \texttt{species}
\citep{stolkerMIRACLESAtmosphericCharacterization2020}.
We adopted the BT-Settl model using solar abundances from CIFIST version \citep{caffauSolarChemicalAbundances2011, allardModelsVerylowmassStars2012, baraffeNewEvolutionaryModels2015}
of HD~142527 A ($T_{\rm eff}$ = 6500 K, $\log{g}$ = 4.0) with a distribution of the surrounding environment, similar template spectrum as used by \citet{christiaensCharacterizationLowmassCompanion2018}. We dereddened it with $A_{V}=0.6 \, mag$ and then scaled it to the radius and distance of the star described in Table \ref{tab:star}. The spectrum was then smoothed with a Gaussian kernel to a spectral resolution of 30 as in IRDIS and IFS observations.

Since the flux of all target sources is measured relative to the central star, it is essential to account for any intrinsic variability of the star. 
HD~142527 A is classified a variable young stellar object with photometric fluctuations observed in a wide range of wavelengths \citep{watsonInternationalVariableStar2006, jayasingheASASSNCatalogueVariable2019, gaiacollaborationGaiaDataRelease2023}.
Its temporal variations in brightness distribution is expected due to the accretion activity as long as its interactions with the disks and the stellar companion B. 
\citet{billerLikelyCloseinLowmass2012} noticed a difference of 0.1--0.3 mag between 2MASS photometry and the \citet{malfaitISOSpectrumYoung1999} photometry. After some years, \citet{mendigutiaSTELLARPARAMETERSACCRETION2014} found its stellar accretion rate increase by a factor $\sim$7 on a timescale from 2 to 5 years. While both \citet{kurtzPulsationPremainsequenceHerbig2001} and \citet{claudiSPHEREDynamicalSpectroscopic2019} reported a 6-day stellar variability period, the peak-to-valley amplitudes differ considerably, with values of 0.13 mag and 0.005 mag in $B$ band, respectively, in different years. These values are not substantial compared to a $10^{-3}-10^{-7}$ contrast between a typical low-mass companion and its star, yet it is necessary to quantify the impacts and include this as a propagated uncertainty in the flux measurements.
Here we compared two synthesized, unsaturated observations of the stellar PSFs taken in the two epochs in 2016. Given that observations are less affected by atmospheric absorption and turbulence in $H$ band, assuming a similar instrumental performance, we expect that any flux difference is attributed primarily to intrinsic stellar variability. We first calculated the integrated flux of the CC in each epoch using the spectra and the IRDIS $H$-band transmission. The difference in flux between epochs was converted to a physical flux uncertainty using a calibrated conversion factor derived from the stellar spectrum. This variability uncertainty was then combined in quadrature with the initial spectral uncertainties, providing a more conservative and realistic error bars.

\begin{figure}[ht]
    \centering
    \subfloat{\includegraphics[width=0.47\textwidth]{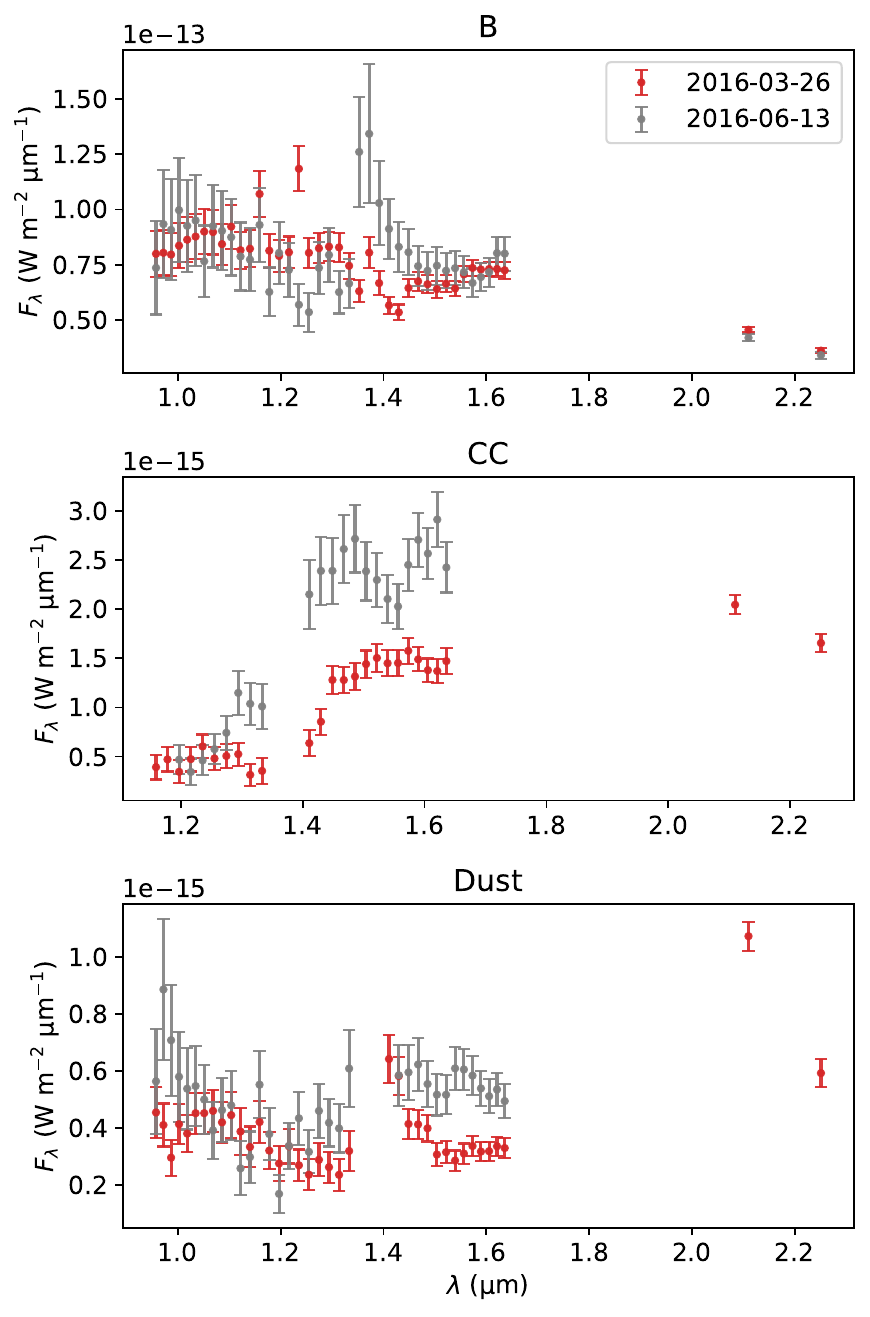}}
    \caption{\label{fig:spectra_all}
    IFS/IRDIS extracted spectrum of HD~142527 B (\textit{top}), the CC (\textit{middle}), and the dust associated with the CC (\textit{bottom}). Error bars are reported with 1$\sigma$ uncertainties.}
\end{figure}

By combining photometry results from different spectral channels, we obtained a low spectral resolution spectrum for the CC and the associated dust signal. These spectra, converted into physical units by multiplying contrasts with the stellar spectrum, are shown in Fig. \ref{fig:spectra_all}.
Generally, the spectra of B between the two 2016 epochs stay within uncertainties, whereas CC and dust spectra exhibit a moderate discrepancy ($\sim$3$\sigma$) toward $H$ band (1.35 $\mu$m onward) between these two epochs. Both objects appear to be redder in June than in March 2016. As the difference is not found in $J$ band, we would argue that the red excess could be due to thermal emission of material along the line of sight. Given that the two sources are marginally resolved, disentangling their individual contributions remains challenging, even with the customized unmixing procedure.

\subsubsection{Astrometry}

To account for the complete error budget of astrometric parameters, we follow the scheme described by \citet[Sect.~4.3.1]{chomezPreparationUnsupervisedMassive2023}. For uncertainties in separations, we adopted the calibration measurements of \citet{maireLessonsLearnedSPHERE2021} and \citet{langloisSPHEREInfraredSurvey2021}, considering image distortion, plate scale, and recentering errors. These were combined quadratically with \texttt{PACO} fitting errors to obtain the total uncertainties. Similarly, uncertainties in position angles were derived from incorporating errors of pupil angle, true north offset, recentering, and \texttt{PACO} estimation errors. In the case of ZIMPOL observations, the same procedure was applied but the calibration measurements are different. It is necessary to correct for instrument north alignment with an uncertainty of 0.5°, as stated by \citet{cugnoSearchAccretingYoung2019}. Details on the observed astrometric positions are given in Table \ref{tab:astrometry}. As a side note, measurements of CC, especially in the 2025 epoch, should be taken with caution because they were retrieved by assuming the detection is point-like, while it is not the case due to the material around CC.

\subsection{HD~142527 B}
\label{B}
\subsubsection{Updates to photometry}
\label{B.1}

The spectral analysis of HD~142527 B has been widely investigated since its detection. Nevertheless, there are significant disagreements between the results across studies. \citet{christiaensCharacterizationLowmassCompanion2018} note a $>$3$\sigma$–discrepancy between measurements using VLT/SINFONI and sparse aperture masking (SAM) in $H$- and $K$- bands in \citet{lacourMdwarfStarTransition2016}, while this difference is within 2$\sigma$ in \citet{billerLikelyCloseinLowmass2012}. Difficulties are also found by \citet{claudiSPHEREDynamicalSpectroscopic2019}, where most data are reused in our study. In their work, HD~142527's spectrum showed a considerable variability across different epochs in $YJH$ band, affecting the fitting reliability. Also reanalyzing some sets of data from  \citet{claudiSPHEREDynamicalSpectroscopic2019}, together with a new SAM observation, \citet{stolkerSearchingLowmassCompanions2023} fit the spectrum of HD~142527 well by inflating the uncertainty scaling and accounting for systematic correlated noise. However, the photometric values extracted from these observations, as others using SAM technique, are still remarkably lower than that of VLT/SINFONI \citep{christiaensCharacterizationLowmassCompanion2018}. While this may attribute to a systematic error, it is also important to note that this companion is not only actively accreting but also embedded in a dusty environment close to the young star, complicating the disentanglement of different contributing components \citep{claudiSPHEREDynamicalSpectroscopic2019}.

\begin{figure*}[ht]
    \centering
    \includegraphics[width=0.8\hsize]{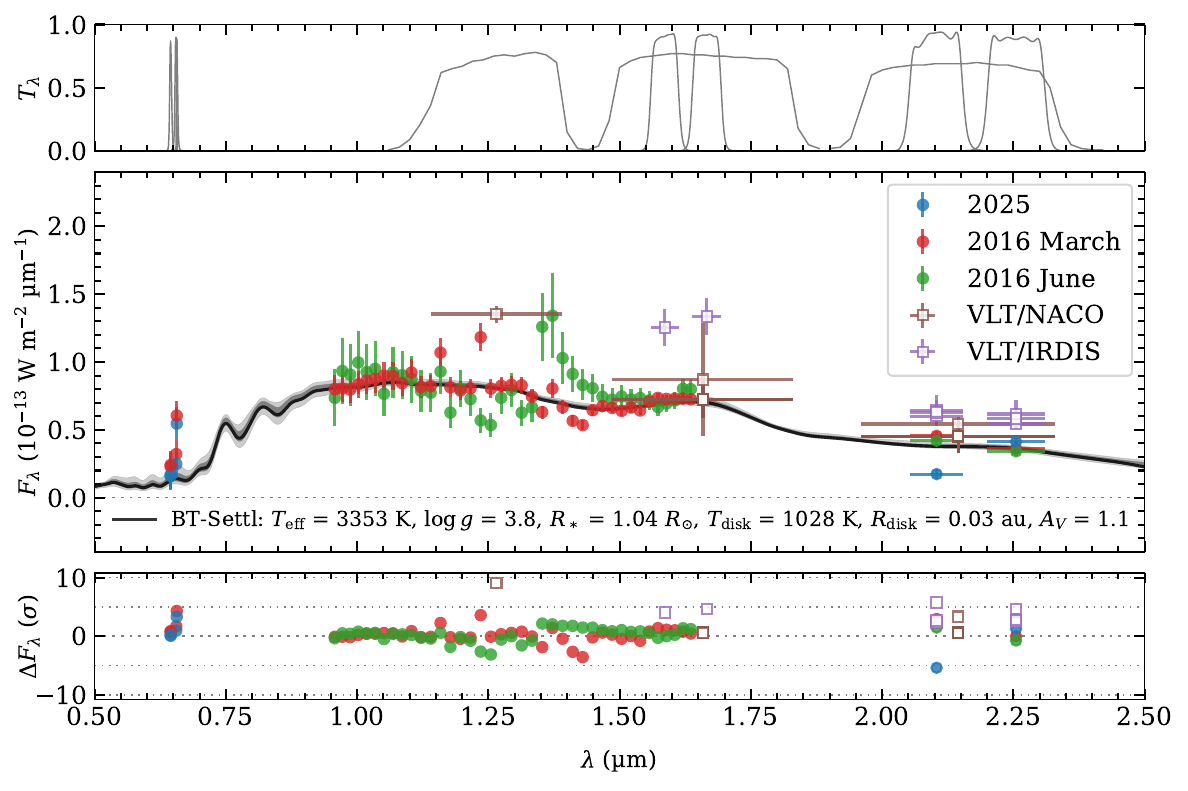}
    \caption{Best-fit models to the SED of the dwarf companion HD~142527 B. Error bars in the $x$-axis of IRDIS and NACO measurements are the full width at half maximum of the corresponding filter transmissions. Reference measurements were extracted from the \texttt{species} library (\citet{lacourMdwarfStarTransition2016} for VLT/NACO and \citet{stolkerSearchingLowmassCompanions2023} for VLT/IRDIS).}
    \label{fig:sed_fit_B}
\end{figure*}

Understanding this challenge, our study adds new insights on the spectrum of the low-mass companion HD~142527 B. We fit the observed low-resolution spectrum with synthetic spectra using \texttt{MultiNest} \citep{ferozMultiNestEfficientRobust2009, buchnerXraySpectralModelling2014}, a multimodal nested sampling algorithm that is integrated in \texttt{species}. Considering uniform distributions for all the prior including effective temperature in range of 3000--4000 K, radius 10--30 $R_\text{Jup}$, interstellar extinction 0.3--1.2, disk radius 15--1000 $R_\text{Jup}$ and disk temperature 900--2000 K, we used 1000 live points to fit the spectrum of B across two epochs in ZIMPOL and two best NIR epochs in 2016. Figure \ref{fig:sed_fit_B} shows best-epoch retrieved spectra of the companion star, together with the values in previous studies \citep{lacourMdwarfStarTransition2016, stolkerSearchingLowmassCompanions2023}. We found best-fit radius of $R = 1.04 \pm 0.01 \, R_\odot$ and  $M= 0.24^{+0.23}_{-0.09} \, M_\odot$ for HD~142527 B. Details of the posterior distribution of the fit can be consulted in Fig. \ref{fig:app_phot_B}.

We also converted the flux contrast of HD~142527 B in H$\alpha$ into magnitude to facilitate the comparison with values obtained in the literature. For the 2016 epoch, the contrast magnitude between B and the star is 5.94 ± 0.04 mag in $N\_Ha$, 6.63 ± 0.08 mag in $B\_Ha$ and 7.00 ± 0.01 mag in the continuum filter. The corresponding values are 6.05 ± 0.12 mag, 6.91 ± 0.14 mag and 7.39 ± 0.02 mag, respectively, for the second epoch in 2025. Compared to contrasts acquired by \citet{cugnoSearchAccretingYoung2019} using the principal component analysis approach, \texttt{PACO} gives refined values with consistency for the same data but toward the 1$\sigma$ lower ends of these estimates, inferring slightly enhanced brightness in all the filters. We also noticed a drop of contrasts across the three filters in the 2025 epoch, but while it is within uncertainties in line H$\alpha$, a more significant change is found in the continuum band ($\sim$0.4 mag). This decrease therefore suggests an intrinsic variability of the companion B rather than a systematic offset.
In addition, our line H$\alpha$ values are in accordance with the ones from MagAO’s VisAO observations at the Magellan Telescope reported by \citet{closeDiscoveryHaEmission2014}, which are 6.33 ± 0.20 mag, but our 2016 continuum contrast is lower than their value, 7.50 ± 0.25 mag. However, we have consistent results with \citet[erratum]{balmerImprovedOrbitalConstraints2022, balmerErratumImprovedOrbital2023}.

\subsubsection{Accretion rate}

Using the H$\alpha$ observations, we calculated the accreting mass rate estimates for HD~142527 B. We adopt the star fluxes obtained by \citet{cugnoSearchAccretingYoung2019}, in which the star extinction and accretion were already taken into account. Considering that there is a variability of 0.09 mag in $R$ band \citep{claudiSPHEREDynamicalSpectroscopic2019}, in order to use the same set of values for star fluxes for both observations in the two epochs, we propagate this into uncertainties of fluxes for all the three bands and show them in Table \ref{tab:halpha}.

\begin{table*}[!ht]
    \caption{\label{tab:halpha}HD~142527 B's H$\alpha$ estimates of flux, luminosity, and accretion rate in different filters.}
    \centering
    \begin{tabular}{llcccccc} 
    \hline \hline
        \multirow{2}{*}{Parameter} & \multirow{2}{*}{Unit} & \multicolumn{3}{c}{30 March 2016} & \multicolumn{3}{c}{27 April 2025} \\
        ~ & ~ & $N\_Ha$ & $B\_Ha$ & $CntHa$ & $N\_Ha$ & $B\_Ha$ & $CntHa$ \\ 
        \hline
        HD~142527 A flux & $10^{-14} \, \text{W\,m}^{-2}$ & {3.0 ± 1.1} & {9.7 ± 1.6} & {6.1 ± 0.7} & {3.0 ± 1.1} & {9.7 ± 1.6} & {6.1 ± 0.7} \\ 
        Observed flux & $10^{-17} \, \text{W\,m}^{-2}$ & {12.6 ± 4.7} & {21.7 ± 3.9} & {9.7 ± 2.1} & {11.4 ± 4.4} & {16.7 ± 3.5} & {6.8 ± 1.6} \\ 
        H$\alpha$ line flux & $10^{-17}\, \text{W\,m}^{-2}$ & 10.3 ± 4.7 & 9.0 ± 4.4 & ... & 9.8 ± 4.4 & 7.9 ± 3.8 & ...\\ 
        H$\alpha$ luminosity & $10^{-5} \, \text{L}_\odot$ & 8.2 ± 3.8 & 7.2 ± 3.5 & ... & 7.7 ± 3.5 & 6.3 ± 3.1 & ...\\
        Accretion luminosity & $10^{-4} \, \text{L}_\odot$ & 15.1 ± 4.4 & 12.8 ± 3.5 & ... & 14.1 ± 4.0 & 10.9 ± 2.7 & ...\\
        Accretion rate & $10^{-10} \, \text{M}_\odot\, \text{yr}^{-1}$ & 2.6 ± 1.2 & 2.2 ± 1.0 & ... & 2.4 ± 1.1 & 1.9 ± 0.8 & ...\\
        \hline
    \end{tabular}
\end{table*}

To calculate HD~142527 B's fluxes that contain both line and continuum contributions, we multiply the contrasts obtained above by the star fluxes. Then, by subtracting these fluxes to the corresponding continuum amounts normalized to H$\alpha$ line bandwidths, we obtain the emission only from the line filters, $F^{\rm{line}}_{N\_Ha}$ and $F^{\rm{line}}_{B\_Ha}$.
The next step is to convert these values into H$\alpha$ luminosity using the relationship $L_{\rm H\alpha} = 4\pi D^{2} \times F_{\rm H\alpha}$, where $D$ is the distance to the system as shown in Table \ref{tab:star}, which yields $L^{\rm{line}}_{N\_Ha} = 8.2 \pm 3.8 \,  L_\odot$ and $L^{\rm{line}}_{B\_Ha} = 7.2 \pm 3.5 \,  L_\odot$ for the 2016 epoch and $L^{\rm{line}}_{N\_Ha} = 7.7 \pm 3.5 \,  L_\odot$ and $L^{\rm{line}}_{B\_Ha} = 6.3 \pm 3.1 \,  L_\odot$ for the 2025 epoch. 
As a side note, for all steps in this analysis, absolute units are used in the calculations and they are converted to solar units solely for presentation purposes.

The empirical relationship between the H$\alpha$ line luminosity and accretion luminosity of young stars and substellar objects is linear \citep{rigliacoXshooterSpectroscopyYoung2012}, and it takes the form
\begin{equation}
    \label{eq:rigliaco}
    \log\left(L_\text{acc}\right) = b + a \times \log\left(L^{\rm{line}}_{\rm H\alpha}\right),
\end{equation}
where $a$ and $b$ are fit from observations depending on different accretion models. We assumed that $a$ = 1.27 ± 0.08 and $b$ = 2.37 ± 0.19, following recent measurements of a wide range of objects across multiple star-forming regions \citep{fiorellinoPENELLOPEVIIRevisiting2025}. 
Then, we got $L^\text{acc}_{N\_Ha}$ for the narrowband filter and $L^\text{acc}_{B\_Ha}$ for the broadband filter.

Afterward, following \citet{gullbringDiskAccretionRates1998}, the accretion rate was estimated as
\begin{equation}
    \label{eq:gullbring}
    \dot{M}_{\text{acc}} = \left(1 - \frac{R_*}{R_{\text{in}}}\right)^{-1} \frac{L_{\text{acc}} R_*}{G M_*} \sim 1.25 \frac{L_{\text{acc}} R_*}{G M_*}, 
\end{equation}
where $R_*$ and  $M_*$ represent radius and mass of the accreting body, in this case HD~142527 B. Same as in \citet{rigliacoXshooterSpectroscopyYoung2012}, we assumed that the accreting disk is truncated at $R_\text{in} \sim 5R_*$, typical for pre-main-sequence stars, resulting in factor $\left(1 - \frac{R_*}{R_{\text{in}}}\right)^{-1}$ being 1.25.

With $R_* = 1.04 \pm 0.01 \, R_\odot$ and $M_* = 0.24^{+0.23}_{-0.09} \, M_\odot$ from our previous spectrum fit, we finally obtained $\dot{M}_{\text{acc}} = 2.6 \pm 1.2 \, \times 10^{-10} \, \text{M}_\odot\, \text{yr}^{-1}$ and $2.4 \pm 1.1 \, \times 10^{-10} \, \text{M}_\odot\, \text{yr}^{-1}$ from narrow H$\alpha$ band in 2016 and 2025, respectively. Correspondingly, the values are $\dot{M}_{\text{acc}} = 2.2 \pm 1.0 \, \times 10^{-10} \, \text{M}_\odot\, \text{yr}^{-1}$ and $1.9 \pm 0.8 \, \times 10^{-10} \, \text{M}_\odot\, \text{yr}^{-1}$ from broadband H$\alpha$ filter.

Although our values for the 2016 dataset generally agree with those from \citet{cugnoSearchAccretingYoung2019} in the broadband filter, there is a deviation of almost 1$\sigma$ in the accretion rate calculated from the narrow H$\alpha$ band. Though we adopted their measurements of star fluxes, we remark that our contrast measurements, as mentioned above, are a little different. We did not find H$\alpha$ flux loss in narrowband filter as reported in their paper, and as a result, the obtained accretion rates are higher in the narrowband filter than in the broadband one. These results align well with theoretical and observational expectations, where narrowband filters isolate the emission line more effectively than broadband data, reducing continuum dilution and enhancing S/N for accretion diagnostics. With \texttt{PACO}, we also addressed the problem of the PSF shape variation between filters, which cause the residual errors in subtraction mentioned by \citet{cugnoSearchAccretingYoung2019}. Therefore, we argue that our measurements are more reliable in this case.
In addition, we also used another set of slope and intercept, $a$ and $b$, in Eq. \eqref{eq:rigliaco}, parameters that were fit from an updated and larger sample in various sky regions while taking a multi-column accretion shock model and extinction into account, compared with the parameters from \citet{rigliacoXshooterSpectroscopyYoung2012}. Hence, the values are more realistic and statistically robust. Moreover, the mass and radius variations of B are also factors that fluctuate the accretion rate estimates, as argued by \citet{balmerImprovedOrbitalConstraints2022}. 
Our accretion rate measurements are also marginally lower than those using MagAO/VisAO H$\alpha$ filters by \citet[erratum]{balmerImprovedOrbitalConstraints2022, balmerErratumImprovedOrbital2023} and possibly of \citet{closeDiscoveryHaEmission2014}, considering conservative uncertainties. 
This discrepancy, apart from being due to the difference in flux calibration, the selection of coefficients $a$ and $b$, as well as radius and mass of B, as previously noted, could also result from an undetected accretion variability of HD 142527 B. However, since our results are within uncertainties, there is not any evident accretion variability observed. The companion's H$\alpha$ accretion had been suggested to vary by \cite{balmerImprovedOrbitalConstraints2022}, but it was later recalculated and confirmed as not being detected \citep{balmerErratumImprovedOrbital2023, folletteGiantAccretingProtoplanet2023}.

\subsubsection{Updates to astrometry}
\label{subsec:B_orbfit}

\begin{figure}[ht]
    \centering
    \includegraphics[width=\hsize]{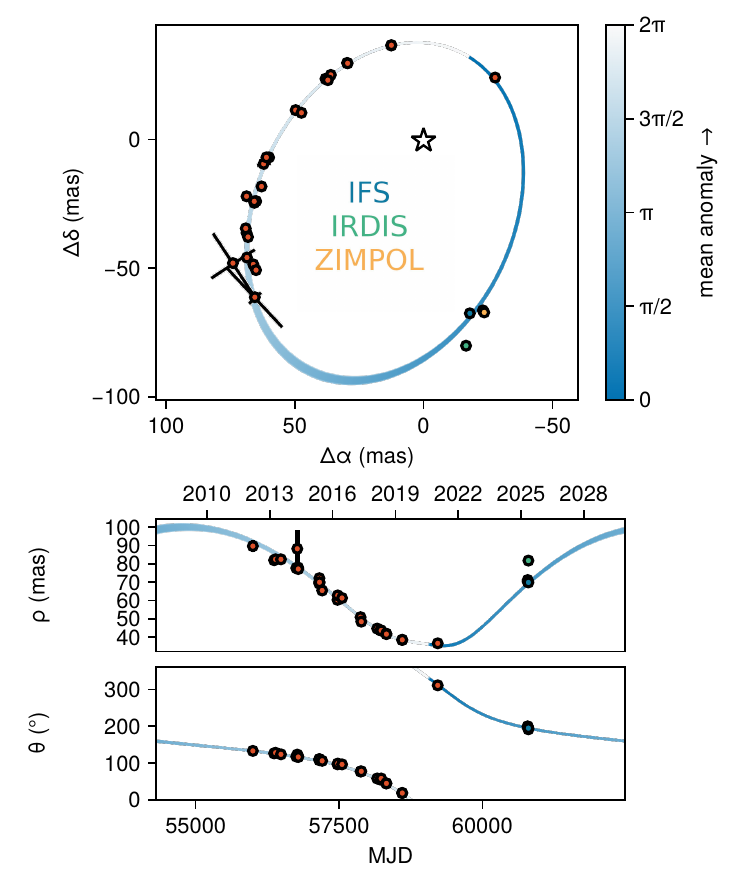}
    \caption{\label{fig:B_astrom}Orbit fit of HD~142527 B with 100 orbits of parameters randomly drawn from the posterior distribution in $\alpha/\delta$ (\textit{top}) and $\rho/\theta$ (\textit{bottom}). Recorded positions are overlaid. Red data points are measurements from the literature, all adopted from \citet{nowakOrbitHD1425272024} in $\alpha/\delta$ format. Blue, green and yellow data points are from ZIMPOL, IFS, and IRDIS observations of this work, respectively. 
    }
\end{figure}

The 2025 observations also offered additional astrometric measurements of HD~142527 B, marking astrometry monitoring during more than half of its orbital period \citep[and references therein]{nowakOrbitHD1425272024}. Figure \ref{fig:B_astrom} shows relative astrometric measurements of 26 points in previous works (red) and three points from our study in the 2025 epoch (blue, green, orange). We implemented the orbit fit using \texttt{Octofitter} \citep{thompsonOctofitterFastFlexible2023} using Hamiltonian Monte Carlo with the No U-Turn Sampler \citep{xuAdvancedHMCjlRobustModular2020} with 1000 iterations, then plotted the probability distributions with  \texttt{PairPlot.jl}\footnote{\url{https://sefffal.github.io/PairPlots.jl/dev}}. We adopted the system's parallax, total mass, as well as B's semi-major axis and eccentricity from the best-fit solution in \citet{nowakOrbitHD1425272024} and sampled these parameters in normal distribution with 3$\sigma$ as priors.

Thanks to the new 2025 observations, we are able to put new constraints on the orbital properties of HD~142527 B (Appendix \ref{sec:app_orbit_B}). Our best-fit parameters are of excellent consistency to the ones reported by \citet{nowakOrbitHD1425272024}, but with smaller uncertainties. B is best fit by a companion of a semi-major axis of $10.67^{+0.17}_{-0.18}$ au, an eccentricity of 0.4652 ± 0.0023, an inclination of $145.32^{\circ}\,^{+0.22}_{-0.19}$, with a total mass for the system of $2.41^{+0.11}_{-0.12}$ $M_\odot$. From the best-fit parameters, the derived period is evaluated as $22.44_{-0.12}^{+0.13}$ years.

\subsection{The candidate companion}
\label{CC}
\subsubsection{Photometry}

\begin{figure*}[ht]
    \centering
    \subfloat{\includegraphics[width=0.455\hsize]{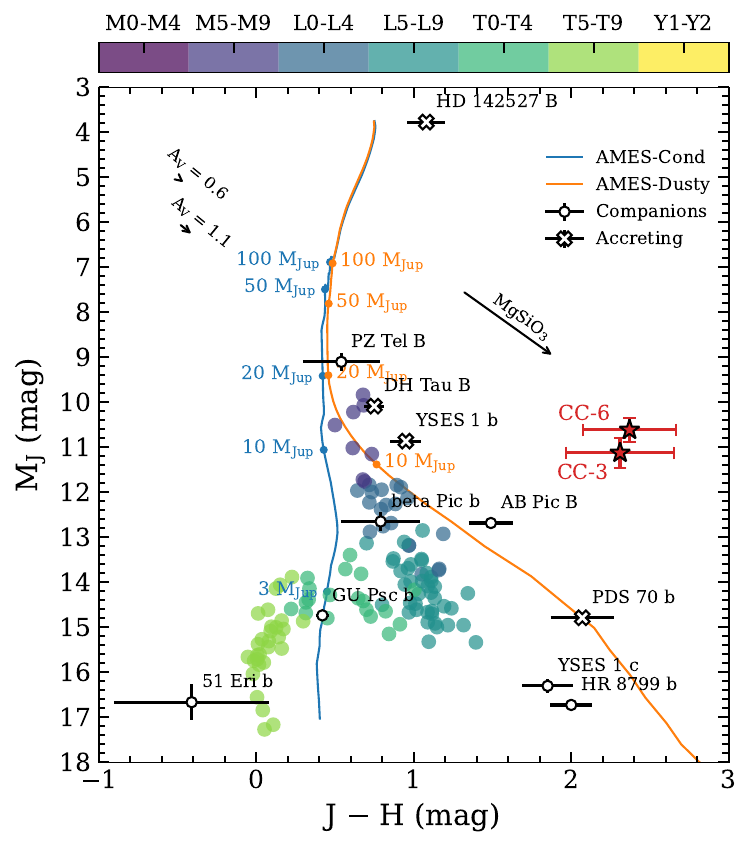}}
    \subfloat{\includegraphics[width=0.45\hsize]{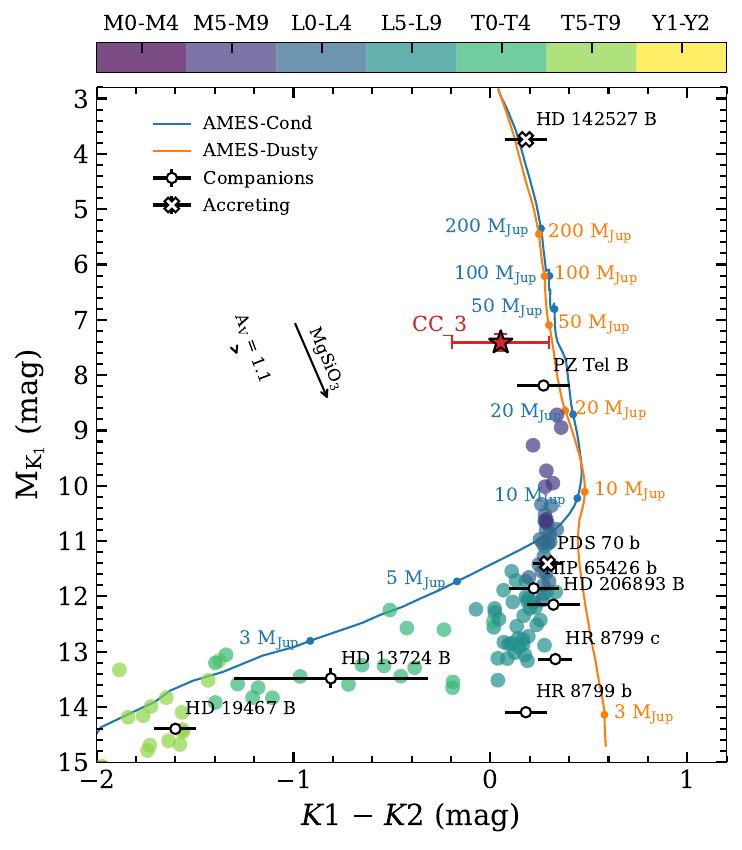}}
    \caption{\label{fig:cmd}Color-magnitude diagram of some known companions with 1$\sigma$ errors. ``CC\_3'' and ``CC\_6'' mark the position of the CC within 3$\sigma$ errors in March and June 2016. Blue and orange curves represent isochrones of different evolution models at the age of 5 Myr. Reddening vectors point to the directions and magnitudes of interstellar reddening and dust reddening as explained in the text.}
    
\end{figure*}

Table \ref{tab:contrast} reports the contrast estimated in different filters of the candidate companion CC in the two 2016 epochs.
The photometric analysis places the CC in the color-magnitude diagram (CMD), as shown in Fig. \ref{fig:cmd}. Spectra for low-mass objects are from the SpeXPrism Spectral Library \citep{burgasserSpeXPrismLibrary2014}. 
The photometry of the reference companions was extracted from the database of \texttt{species}.
\begin{table}[ht]
    \caption{\label{tab:contrast}Flux contrast and absolute magnitudes of the CC.}
    \centering
    \begin{tabular}{cccc}
        \hline \hline
        \multirow{2}{*}{Filter} & Contrast & App. magnitude & Abs. magnitude \\
        ~ & ($\times 10^{-5}$) & $m_\lambda$ (mag) & $M_\lambda$ (mag) \\ \hline 
        \multicolumn{4}{c}{26 March 2016} \\ \hline
        $J$ & 6.9 ± 0.4  
        & 17.13 ± 0.11
        & 11.12 ± 0.11
        \\ 
        $H$ & 25.1 ± 0.5 
        & 14.82 ± 0.03 
        & 8.81 ± 0.03 
        \\
        $K1$ & 57.5 ± 2.1 
        & 13.42 ± 0.05 
        & 7.41  ± 0.05 
        \\ 
        $K2$ & 51.0 ± 2.1 
        & 13.37 ± 0.06 
        & 7.36  ± 0.06 
        \\ 
        \hline
        \multicolumn{4}{c}{13 June 2016} \\ \hline
        $J$ & 11.5 ± 0.9  
        & 16.51 ± 0.11
        & 10.50 ± 0.11
        \\
        $H$ & 46.1 ± 1.4  
        & 14.25 ± 0.04 
        & 8.24 ± 0.04
        \\
        \hline
    \end{tabular}
\end{table}
We considered extinction and reddening, i.e., an object appearing fainter and redder than it is in reality. This effect is represented in the CMD by reddening vectors, which point to the direction in which reddening and extinction influence the object's color and magnitude. Apart from interstellar reddening that is due to the interstellar medium, dust reddening should be considered if there is additional dust that is not accounted for in atmospheric models or dust in the line of sight. Then, the intrinsic color and magnitude of the object are estimated by sliding the position back along the vector. We include $A_V = 0.6$ mag and $A_V = 1.1$ mag, the extinction of the central star A and that of the companion B from the best fit in Sect. \ref{B.1}, to show that $A_V$ of CC is probably way higher. Therefore, we consider a power-law distribution of crystalline enstatite (MgSiO\textsubscript{3}) grains (see the SED fit at the end of this section). In
the CMD of $J$-band magnitude and $J-H$ color, CC appears remarkably red compared to known companions and evolutionary models. 
However, as mentioned earlier, the spectra in $H$ band of CC could be affected by dust thermal emission along the line of sight, so it is likely that CC is subject to strong reddening, pushing it away from the theoretical isochrone curves. In this case, a power-law distribution of enstatite could explain its redness, pulling it closer to other known companions like DH Tau B or YSES 1b. On the other side,
in the $K1-K12$ CMD, CC is placed next to both the isochrones, implying a mass between 20 and 50 M\textsubscript{Jup}. Nevertheless, the enstatite dust does not explain its color and magnitude anymore. Several possibilities could account for this discrepancy, including that the actual dust population or distribution differs from our assumptions, or that there is additional blue excess due to CC accreting activities that we have yet to detect.

Assuming the CC is a point source, we performed spectrum fitting to explore its nature. Given the discrepancy between the two best epochs, we considered fitting only the SED of the most reliable epoch, on 26 March 2016, where the signal from CC is well separated from the surrounding contamination. However, located inside the extent of the inner disk, it is conceivable for the CC to be linked with dust, either in its atmosphere or around it, in a circumplanetary or circumstellar disk. Therefore, it is not unexpected if existing atmospheric models fails to reproduce its spectrum. As a consequent, instead of focusing on finding the exact parameters that describe the CC, we aim just to have a first overview which may give a clue about its nature.

\begin{figure}[ht]
    \centering
    \includegraphics[width=\columnwidth]{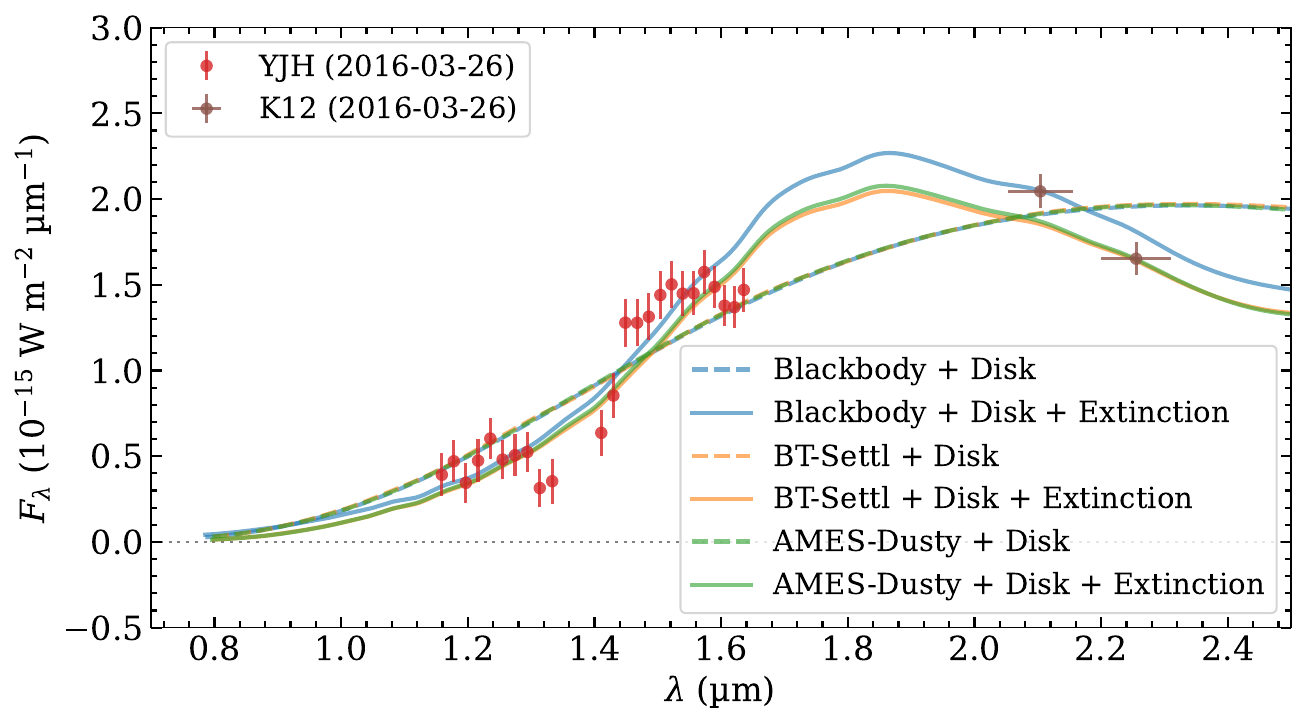}
    \caption{Best-fit models to the SED of the CC. Error bars in the $x$-axis of IRDIS measurements are the full width at half maximum of the $K1$ and $K2$ filter transmissions.}
    \label{fig:sed_fit_CC}
\end{figure}

\begin{table*}[ht]
    \caption{\label{tab:bestfit_params} Fit parameters and derived luminosity and mass of the CC for different models.}
    \centering
    \begin{tabular}{lcccccc}
    \hline \hline
        \multirow{2}{*}{Parameters} & \multicolumn{2}{c}{Blackbody} & \multicolumn{2}{c}{BT-Settl} & \multicolumn{2}{c}{AMES-Dusty}\\ 
        ~ & Disk & Disk + Extinction & Disk & Disk + Extinction & Disk & Disk + Extinction \\ \hline
        $T\textsubscript{eff}$ (K) & $1448^{+225}_{-141}$ & $2133^{+470}_{-554}$ & $1395^{+247}_{-110}$ & $1653^{+689}_{-374}$ & $1432^{+135}_{-123}$ & $1528^{+641}_{-265}$ \\
        $\log{g}$  & ... & ... & $4.43^{+0.78}_{-0.75}$ & $4.43^{+0.57}_{-0.56}$ & $4.44^{+0.66}_{-0.53}$ & $4.37^{+0.59}_{-0.49}$ \\ 
        R (R\textsubscript{Jup}) & $1.36^{+1.05}_{-0.65}$ & $1.63^{+0.73}_{-0.70}$ & $0.80^{+0.53}_{-0.21}$ & $1.09^{+0.83}_{-0.39}$ & $1.16^{+0.91}_{-0.47}$ & $1.20^{+1.06}_{-0.49}$ \\ 
        $\pi$ (mas) & $6.28 \pm 0.02$ & $6.28 \pm 0.02$ & $6.28 \pm 0.02$ & $6.28 \pm 0.02$ & $6.28 \pm 0.02$ & $6.28 \pm 0.02$\\
        $T\textsubscript{disk}$ (K) & $1237^{+18}_{-20}$ & $1064^{+64}_{-72}$ & $1236^{+18}_{-19}$ & $1088^{+47}_{-53}$ & $1245^{+18}_{-19}$ & $1098^{+43}_{-52}$ \\ 
        $R\textsubscript{disk}$ (R\textsubscript{Jup}) & $15.66^{+0.88}_{-0.78}$ & $59.69^{+30.69}_{-16.92}$ & $15.84^{+0.84}_{-0.79}$ & $53.96^{+34.20}_{-12.14}$ & $15.45^{+0.81}_{-0.83}$ & $52.90^{+30.99}_{-11.11}$ \\ 
        $\log a_{max}$ ($\mu$m) & ... & $0.32 \pm 0.02$ & ... & $0.32 \pm 0.02$ & ... & $0.32 \pm 0.02$ \\ 
        $\beta\textsubscript{dust}$ & ... & $5.7^{+3.1}_{-4.3}$ & ... & $4.4 \pm 3.8$ & ... & $4.2^{+3.9}_{-3.6}$ \\ 
        $A_{V}$ & ... & $1.5^{+0.4}_{-0.3}$ & ... & $1.5^{+0.6}_{-0.3}$ & ... & $1.5^{+0.5}_{-0.3}$ \\ \hline
        Derived $\log L_{p}/L_\odot$ &
        $-4.06^{+0.40}_{-0.48}$ 
        & $-3.32^{+0.44}_{-0.64}$ 
        & $-4.54^{+0.34}_{-0.31}$ 
        & $-3.97^{+0.46}_{-0.51}$ 
        & $-4.25^{+0.47}_{-0.42}$ 
        & $-3.97^{+0.44}_{-0.51}$ \\ 
        Derived M ($M\textsubscript{Jup}$) 
        & ...
        & ...
        & $9^{+25}_{-7}$
        & $17^{+26}_{-13}$ 
        & $22^{+29}_{-17}$ 
        & $19^{+27}_{-14}$ \\ 
        \hline
        $\chi^2$ & 3.92 & 2.83 & 4.19 & 2.85 & 4.11 & 2.73 \\ 
        \hline
    \end{tabular}
    \tablefoot{
    The values are the median of 30 samples extracted from the posterior of the fits, with uncertainties representing the 16\textsuperscript{th} and 84\textsuperscript{th} percentiles of these samples.
    }
\end{table*}

First, we fit the SED with a set of atmospheric models: blackbody, BT-Settl \citep{allardModelsVerylowmassStars2012} and
AMES-Dusty \citep{allardLimitingEffectsDust2001}. We chose uniform priors of effective temperature $T\textsubscript{eff}$ = (1000, 3000) K, radius $R$ = (0.5, 3.0) $R\textsubscript{Jup}$ for all the models; for BT-Settl, we set surface gravity $\text{log }g$  = (3.5, 5.5), while for AMES-Dusty, $\text{log }g$ = (3.5, 6.0).
In addition, we added a second blackbody to simulate excess disk emission from the hot environment around CC, parameterized by $T\textsubscript{disk}$ in the range (900, 2000) K and $R\textsubscript{disk}$ between 15.0 and 150.0 $R\textsubscript{Jup}$. Based on the CMD, we imposed a normal distribution for CC's mass prior to be 40 ± 30 $M\textsubscript{Jup}$.

Given the possibility that the CC is shrouded by dust, we considered also the effects of extinction following the extinction law by \citet{cardelliRelationshipInfraredOptical1989}. We did not force the value of $A_{V}$ to be the same as of the host star in Table \ref{tab:star} because there might be a difference in the environment surrounding them. Then, we considered a more general case where dust size distribution is taken into account when estimating the extinction. 
Assuming a distribution of crystalline enstatite grains, which is found to be abundant in atmospheric layers of cool dwarfs \citep{allardLimitingEffectsDust2001, morleyNEGLECTEDCLOUDSDWARF2012} with a homogeneous, spherical structure, the power-law size distribution is defined as $n \propto a^{-\beta\textsubscript{dust}}$ where $a$ is the particle radius and $\beta\textsubscript{dust}$ the power-law exponent \citep[see also][]{mathisSizeDistributionInterstellar1977, wangConstrainingNaturePDS2021}. We fixed the minimum radius to 1 nm and fit the two parameters $a_{\rm{max}}$ and $\beta\textsubscript{dust}$ along with $A_{V}$. Priors are uniform for $\log a_{\rm{max}}$ in the range (--2.0, 1.0), $\beta\textsubscript{dust}$ in range (--10.0, 10.0) and $A_{V}$ in range (0.0, 5.0) mag. The wavelength-dependent extinction cross sections pre-calculated by \citet{mollierePetitRADTRANSPythonRadiative2019} are stored in a grid in \texttt{species} database. 

We implemented the fit using \texttt{Multinest} with 1000 live points. Figure \ref{fig:sed_fit_CC} displays CC spectrum fitting using different models, which are also listed in Table \ref{tab:bestfit_params}. Though with diverse assumptions to address to the red color of CC, it is noticeable that the fits are with a high level of similarities. In general, CC's effective temperature is estimated from 1300 to under 2200 K, with large surface gravity ($\log{g} \sim$ 4.4) and a radius of 0.80 to 1.63 $R\textsubscript{Jup}$.
In the presence of a blackbody disk accounting for excess thermal emission, its radius and temperature change with the absence or presence of extinction with power-low size distribution. Specifically, the disk temperature is approximately 150 K lower and the disk radius is 4 to 5 times larger in case of extinction, compared to the case without. In all fits with extinction, the extinction index is expected to be around 1.5 and the particle radius approximately 0.32 $\mu$m in all models. In the end, different models give different mass estimates, from less than 10 $M\textsubscript{Jup}$ to several tens of $M\textsubscript{Jup}$ due to large uncertainties. We would argue that values of at least 10 $M\textsubscript{Jup}$ are more realistic, given the mass limits of 12--20 $M\textsubscript{Jup}$ (Fig. \ref{fig:contrast_curves}).

\subsubsection{Astrometry and orbital fit}
\label{subsec:CC_orbfit}

\begin{figure}[ht]
    \centering
    \subfloat{\includegraphics[page=1,width=.8\hsize]{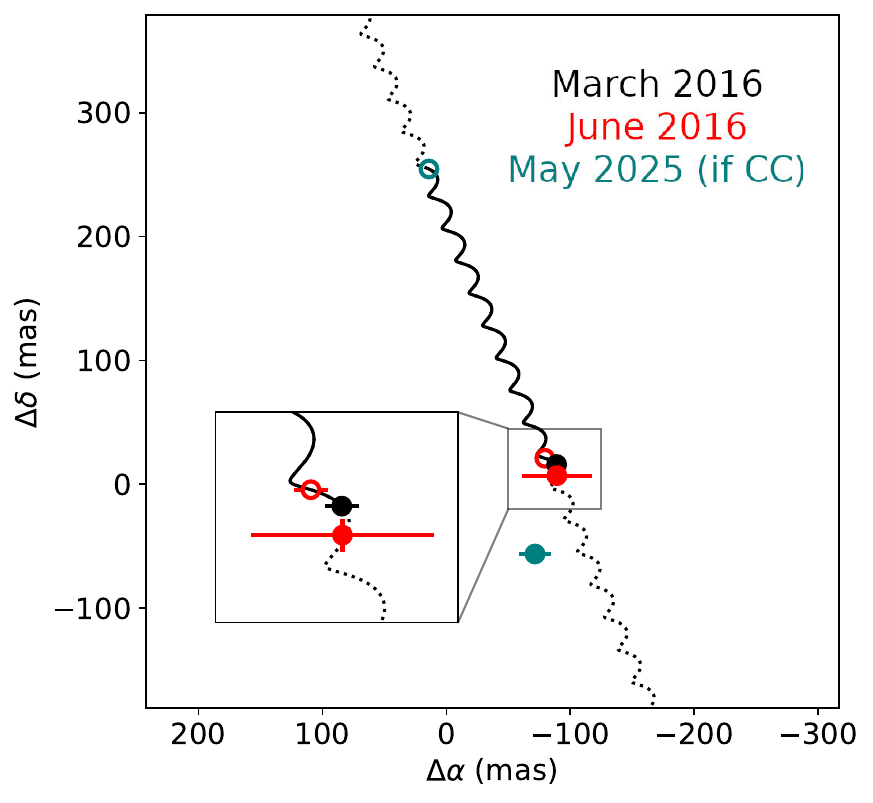}}
    \\
    \subfloat{\includegraphics[page=2,width=.8\hsize]{CC_astrometry_rev_v2}}
    \caption{\label{fig:astrometry}Astrometry displacements of the CC relative to the star in $\alpha/\delta$ (\textit{top}) and $\rho/\theta$ (\textit{bottom}). Color-filled dots represent the detected positions of the CC in the two 2016 epochs and the putative CC signal in 2025, while open dots of the corresponding colors indicate expected positions of the CC if it is a background source. Uncertainties of the CC data points and the expected background motion are both in 1$\sigma$, while the putative the CC position in 2025 is marked with 3$\sigma$ for a more robust comparison. 
    }
\end{figure}

Figure \ref{fig:astrometry} illustrates astrometric displacements of CC recorded in IFS observations in the two epochs in March and June 2016, alongside the suspected 2025 CC signal position.
With just the archived 2016 observations, it is unclear whether CC shares a gravitational association with the host star. The short temporal baseline between the two detections prevents a conclusive common proper motion confirmation. However, CC remains likely incompatible with a background source in the 3$\sigma$ range of position angle, implying it is bound to the host star. 
While the suspected signal in 2025 cannot yet be proven as CC, if it were, its position far from that of an background source would be a clear evidence suggesting its gravitational link to the star.

In addition, we also estimated the probability to find a background contaminant around the central star HD~142527 within CC's detected positions in the two epochs in 2016 and 2025. From the database of the Besançon galactic model \citep{robinSyntheticViewStructure2003}, which provides the stellar density at specific galactic coordinates and apparent magnitudes, adapted for the SPHERE near-infrared survey, we determined the number of stars within a field of 0.1$''$ that exceeds the $H$-band's brightness of CC, i.e., with apparent magnitude less than 14.535, the average $H$-band magnitude of CC in the two epochs given in Table \ref{tab:contrast}. The probability of having at least one background star in this region is found to be very small, at $7.5 \times 10^{-7}$.

Assuming the 2025 signal is CC, we fit its orbit jointly with B using \texttt{Octofitter}, following the method shown in Sect. \ref{subsec:B_orbfit}. We kept all data points as well as prior distribution of B and system parallax as in Fig. \ref{fig:B_astrom} of the previous analysis. Then, CC is added to the system with its positions as shown in Fig. \ref{fig:astrometry}. 
We applied the observable-based priors technique from \citet{oneilImprovingOrbitEstimates2019}, which is optimal for low-phase-coverage orbits. 
To make it compatible, instead of imposing a prior distribution on CC's semi-major axis, we did it on its period, assuming a uniform sample from 25-100 years. The semi-major axis can be derived afterward using Kepler's third law. In addition, we assume a normal distribution of CC's mass $M_{\rm{CC}} = 40 \pm 30 M\textsubscript{Jup}$. Considering the complexity of the orbit fit, we utilized the dedicated Markov chain Monte Carlo scheme for challenging, multimodal posteriors non-reversible parallel tempering \citep{syedNonReversibleParallelTempering2022} in the Julia package \texttt{Pigeons} \citep{surjanovicPigeonsjlDistributedSampling2023} integrated in \texttt{Octofitter}. The orbit fit was implemented using slice sampling \citep{nealSliceSampling2003} explorer in 2\textsuperscript{14} (i.e., 16384) iterations, with two-leg variational parallel tempering of 20 chains between the posterior and the prior and 20 chains between the posterior and a variational reference.
From the best-fit parameters, we also evaluated the mutual inclination between B and CC using the following formula:
\begin{equation}
    \label{eq:mulinc}
    i_{\rm mut} = \arccos\left( \cos i_1 \cos i_2 + \sin i_1 \sin i_2 \cos(\Omega_1 - \Omega_2) \right),
\end{equation}
where $i$ and $\Omega$ indicate the inclination and longitude of ascending nodes, either of B or of CC.

\begin{figure}[ht] 
    \centering
    \includegraphics[width=0.9\hsize]{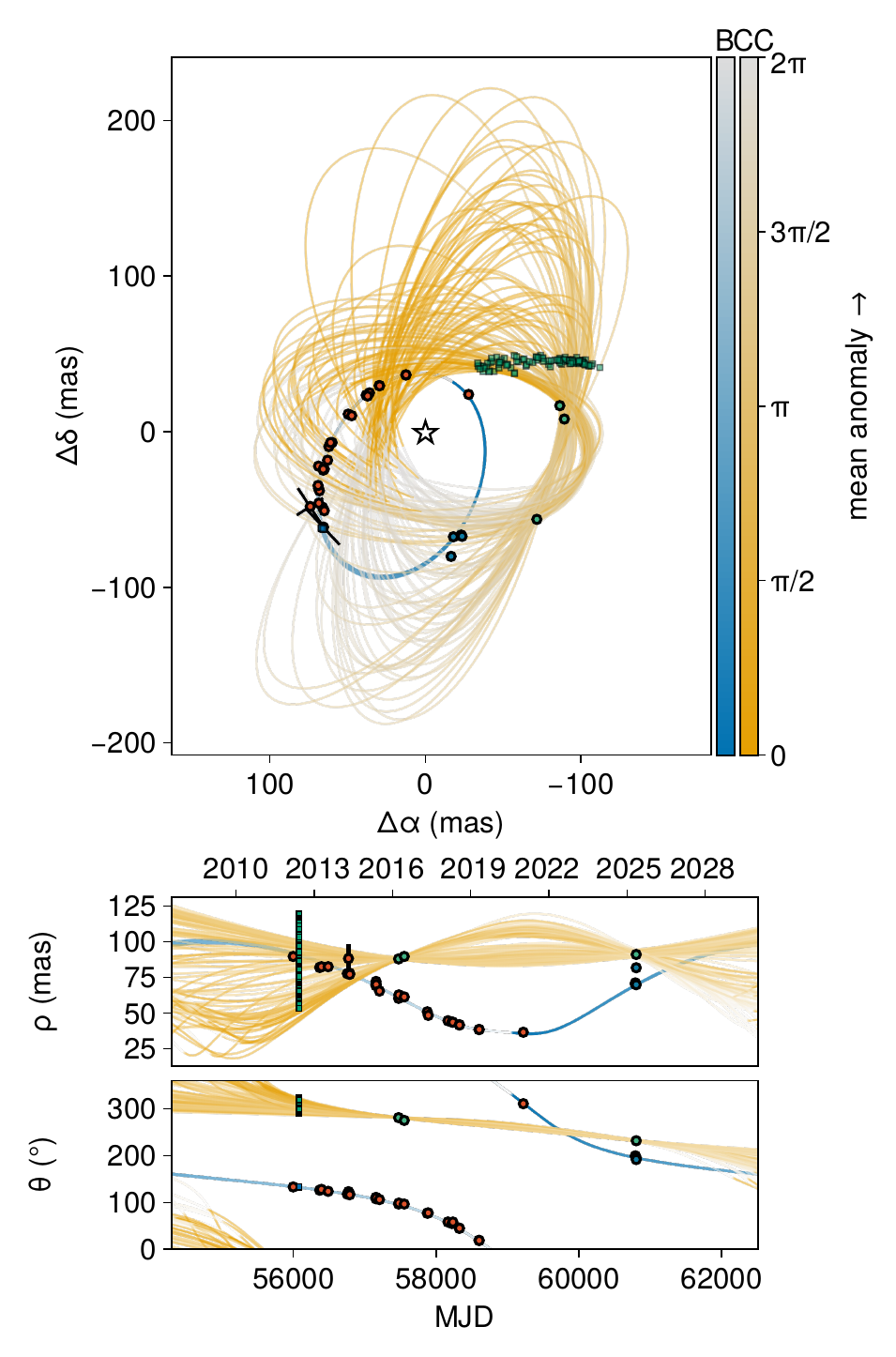}
    \caption{\label{fig:both_orbits}
    Simultaneous orbit fit of both HD~142527 B and CC (assuming third-epoch measurement in 2025) with 100 sample orbits randomly drawn from the posterior distribution. Measurements are displayed in circles, with B's positions in red (reference points) and blue (from this study), the same as in Fig. \ref{fig:B_astrom}. Green circle points indicate the positions of the CC in 2016 and the putative CC signal in 2025. Additionally, square points of corresponding colors indicate projected positions of B and the CC from the orbital fit on 1 June 2012. 
    }
\end{figure}

Figure \ref{fig:both_orbits} displays the simultaneous orbit fit of B and CC, with 100 randomly orbits drawn from the posterior distributions (see also Appendix \ref{sec:app_orbit_all}). The estimated parameters of B do not change significantly compared to those found by fitting only B. However, given the few measurements of the CC's astrometry, its orbital parameters are less constrained, showing mutual correlation. Assuming the 2025 signal, some parameters of CC converge to bimodal distributions, including semi-major axis (one peak at $\sim$ 16 au, another at $\sim$ 27 au), eccentricity (0 and $\sim$ 0.5), and period ($\sim$ 40 and $\sim$ 84 years). From the posterior distribution, it is likely that CC's orbit is more eccentric when its semi-major axis is smaller, as shown by the negative slopes in the correlation between these two parameters. 
Similarly, the double peaks in the distribution of CC's period are consistent with those of the semi-major axis, confirmed by the clear correlation between them.
Using the posterior medians with 68\% uncertainties, the orbit of CC is best fit with semi-major axis $18^{+9}_{-3}$ au, eccentricity $0.5^{+0.1}_{-0.3}$ and inclination $119^{\circ}\,^{+7}_{-9}$. Its mass is calculated to be $41^{+29}_{-23} \, M\textsubscript{Jup}$. The derived period is evaluated as $44^{+40}_{-9}$ years. In addition, the mutual inclination between B and CC is found to be $61^{\circ}\,^{+39}_{-25}$.

In any case, we acknowledge that associating the CC with the 2025 red signal is a strong assumption given the limited astrometric data. Further astrometric measurements are indispensable to confirm whether the CC has a stable orbit around HD~142527 and, in that case, to refine its orbital parameters.

\subsubsection{Dynamical stability analysis}

\begin{figure*}[ht]
    \centering
    \includegraphics[width=0.8\hsize]{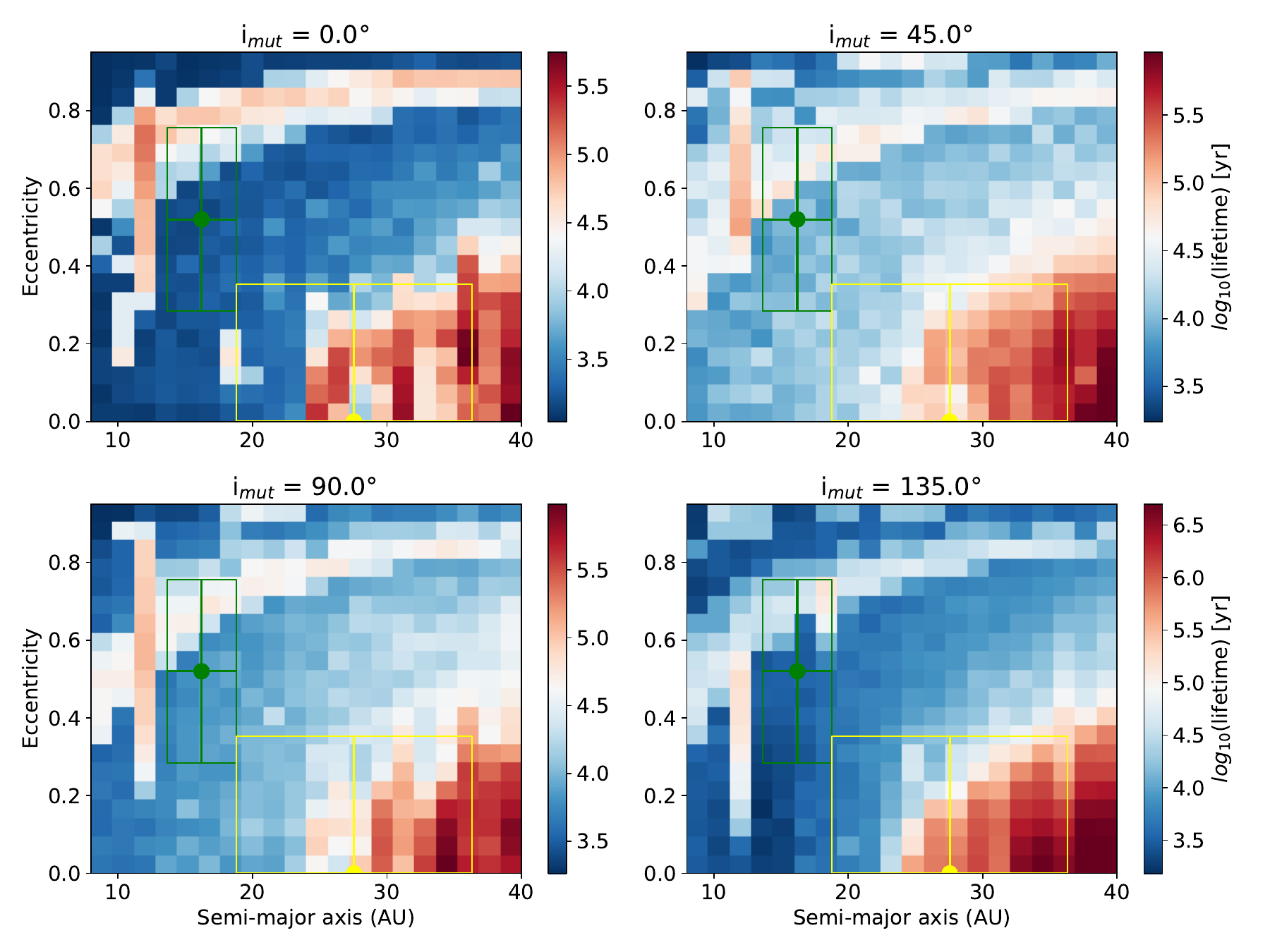}
    \caption{\label{fig:stability}
    Dynamical stability of the CC as a function of its semi-major axis and eccentricity in case of varying mutual inclinations with B, which are 0\textdegree, 45\textdegree, 90\textdegree,\, and 135\textdegree. Square boxes represent the two regions of semi-major axis and eccentricity extracted from their bimodal posterior distributions of the CC's orbit fit in Sect. \ref{subsec:CC_orbfit}.
    }
\end{figure*}

To investigate the dynamical evolution and stability of CC, we ran N-body simulations using the fast and unbiased WHFast integrator \citep{wisdomSymplecticMapsNbody1991, reinWHFASTFastUnbiased2015} in \texttt{REBOUND} \citep{reinREBOUNDOpensourceMultipurpose2012}. Given the lack of information of CC except a crude estimation of its mass, we explored possible orbital configurations in which CC can stay dynamically stable in the system. Assuming a single-point-mass central star and a $\sim 30 \,M\textsubscript{Jup}$ CC mass, we created test systems including the three bodies with their masses: $m_{A} = 2.0 \pm 0.3 \, M_\odot$ (as shown in Table \ref{tab:star}), $m_{B} = 0.24^{+0.23}_{-0.09} \, M_\odot$ from our previous spectrum fit, and $m_{CC} = 0.03 \, M_\odot$. Mass uncertainties ($m_{A}$, $m_{B}$) are fundamental and directly affect gravitational interactions and Hill radii, making them the dominant parameter for stability in hierarchical systems \citep{verasDynamicsTwoMassive2004}, and thus we took them into consideration using uniform sampling across 3$\sigma$ mass range of each body. As B's orbit was well constrained by \citet{nowakOrbitHD1425272024}, we fixed it with $a_{B} = 10.80$ and $e_{B} = 0.47$ as in the article. We ignored the companion's argument of periapsis and mean anomaly, considering that long integration timescales average out the impact of initial orbital phase. By varying CC's semi-major axis in range from 8 to 40 au, eccentricity from 0 to 0.95, argument of periapsis $\omega$ and mean anomaly $M$ each from 0$^\circ$ to 360$^\circ$, we created an initial parameter grid of CC ($a \times e \times \omega \times M$) = ($20 \times 20 \times 5 \times 5$) for each configuration of the binary A-B. Next, these test systems were integrated for 5 Myr, the age of the central star, to record the lifetime of the CC. Specifically, the lifetime is either the maximum, 5 Myr, or the time from the beginning of the integration until the CC collides or is ejected, which is determined when its distance from the barycenter reaches less than 0.1 or is greater than ten times its initial semi-major axis. Subsequently, for each ($a$, $e$) pair, CC's lifetime was averaged across all possible configurations to generate the stability maps.

In addition, mutual inclinations could critically impact triple system dynamics by triggering secular disturbances that affect stability lifetime.
We incorporated this into our simulations by varying the mutual inclinations among (0$^\circ$, 45$^\circ$, 90$^\circ$ and 135$^\circ$), assuming $\Omega_{CC} = 161.51^{\circ}$, the same as $\Omega_{B}$ found by \citet{nowakOrbitHD1425272024}. 
This set of mutual inclinations covers our range of $i_{\rm mut} = 61^{\circ}\,^{+39}_{-25}$ found earlier in CC's orbit fit.

Figure \ref{fig:stability} presents stability maps of the CC where different mutual inclinations with respect to B have been assumed. Across all panels, there are two regions that allow the CC to remain stable for over 100,000 years: (A) the region with a smaller semi-major axis and high eccentricity and (B) the region with a larger semi-major axis and a low eccentricity. These results agree with the orbit fit in Sect. \ref{subsec:CC_orbfit}, where we found bimodal distributions of CC's semi-major axis and eccentricity.

In case (A), the stability favors a narrow strip of $a \sim12$ au and $e$ from 0.3 to 0.8, but at higher eccentricities, the semi-major axis of CC could also fall in a larger range for stability. This likely results from B-CC interactions such as mean motion resonance or secular perturbations like Kozai-Lidov mechanism \citep{kozaiSecularPerturbationsAsteroids1962, lidovEvolutionOrbitsArtificial1962} that allow some particular sets of parameters, even with extremely high eccentricities and retrograde orbits \citep[see, e.g., the review of][]{katzLongTermCyclingKozaiLidov2011}. In this parameter space zone, CC can live up to several hundred thousand years, approximately 10\% of the system age. Though this suggests that CC was observed in a transient stage and will finally be removed, its lifetime is still considerable compared to that of the system.

Conversely, case (B) presents feasible configurations with more circular, distant orbits. Because we detect CC at $\sim 0.09''  \sim 14$ au, this case is possible when the orbits are inclined, which is confirmed in the orbit fit in Sect. \ref{subsec:CC_orbfit}. Across all inspected maps, CC's lifetime is estimated much longer than case (A), up to millions years.

We then compared our results of the CC's orbit fit with its dynamical analysis by overplotting the two distinct ($a$, $e$) regions extracted from the posterior distributions in Sect. \ref{subsec:CC_orbfit} onto Fig. \ref{fig:stability}. The peaks of these parameters were evaluated by fitting a Gaussian model to their histograms, using the full width at half maximum as the uncertainty. It can be seen that within these uncertainties, numerous possible configurations favor the CC's dynamical stability in the system, conforming to either case (A) or (B).

\subsection{The spiral feature}
\label{sec:spiral}

Spirals in circumstellar disks are closely linked to planet formation. They can be caused by various phenomena, such as gravitational interactions with unseen planets triggering density waves in the gas component or temperature variations, or other sources of instability, such as Rossby waves \citep{kleyPlanetDiskInteractionOrbital2012}.

\begin{figure}[ht] 
    \centering
    \subfloat{\includegraphics[width=\hsize]{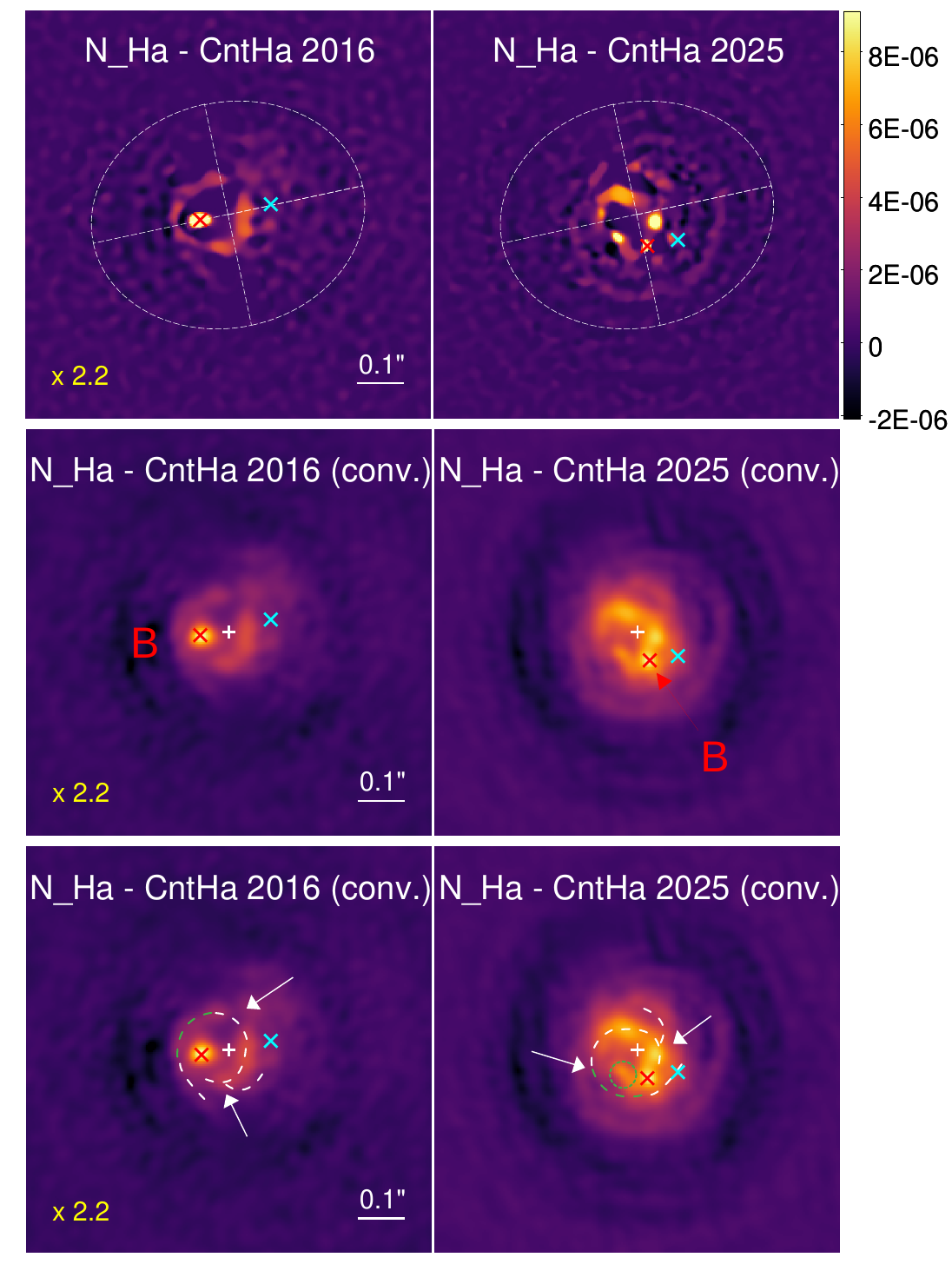}}
    \caption{\label{fig:spiral_line}\textit{Top}: Deconvolved ZIMPOL H$\alpha$ line flux distribution of HD 142527 (continuum-subtracted narrowband images), reconstructed by \texttt{REXPACO}. Both panels use the same linear scale in flux contrast and are overlaid with the same ellipse from Sect. \ref{sec:results}. B (red $\times$) is saturated to enhance faint structures, while the CC and the suspected 2025 signal are marked by a cyan $\times$. Since the signal is brighter in 2025 (see text), the intensity scale factor to ensure displaying uniformity is marked in the 2016 images' corner.
    \textit{Middle}: Images in the top panels convolved again with the instrument's PSF, shown in square-root scale to enhance the display of faint structures.
    \textit{Bottom}: Same as the middle panels but with annotations overlaid for faint features that possibly co-rotate with B (white, found in both epochs; green, found in one epoch only). From 2016 to 2025, these annotations are rotated 260$^\circ$ clockwise to follow B's movement.
}
\end{figure}

\begin{figure}[ht] 
    \centering
    \subfloat{\includegraphics[width=\hsize]{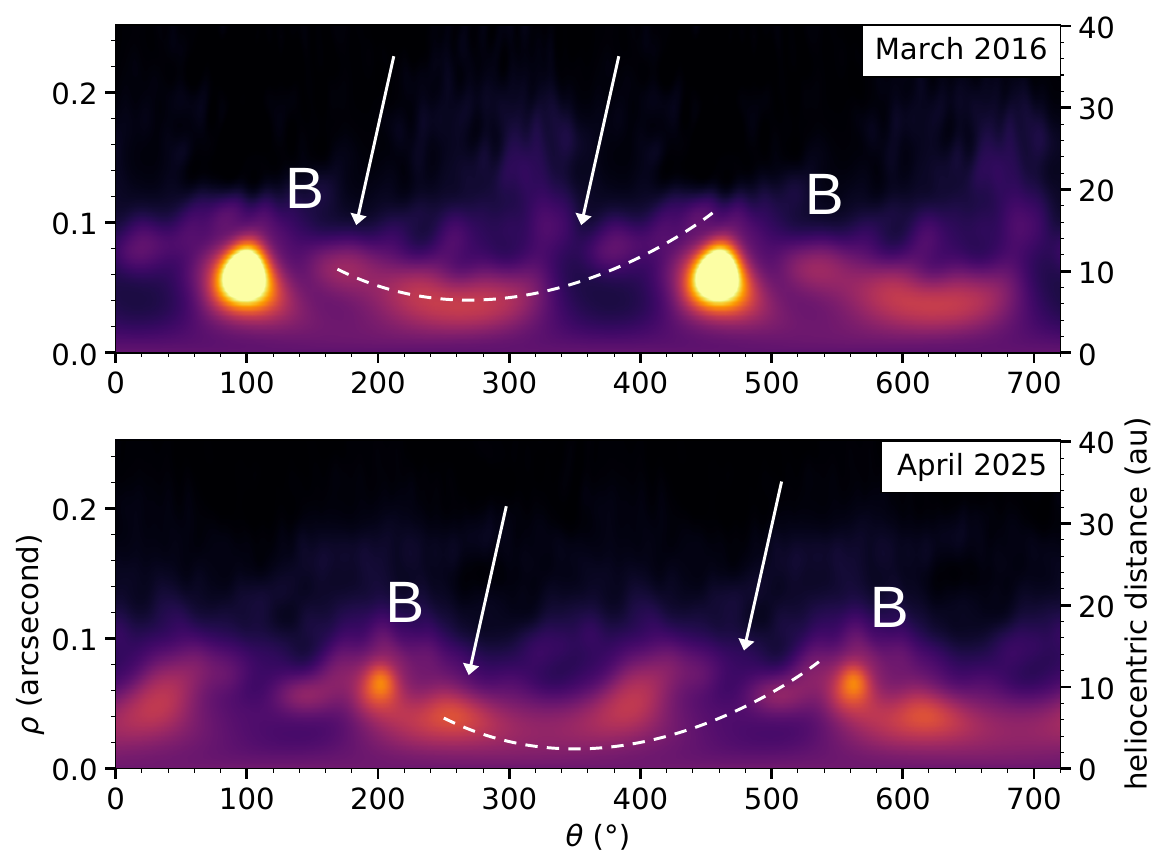}}
    \caption{\label{fig:spiral_deprojected}
    Bottom images of (a) deprojected to $\theta/\rho$.
    Dashed lines and white arrows point to positions similar to those in (a), suggesting hints of co-rotation of the features with B.
}
\end{figure}

HD~142527 has been known for its outer disk with spectacular spiral features that have been observed in both scattered light and thermal emission \citep{casassusDYNAMICALLYDISRUPTEDGAP2012, avenhausStructuresProtoplanetaryDisk2014, christiaensSPIRALARMSDISK2014}.
In this study, we detect a spiral-like feature in the immediate vicinity of HD~142527, revealed through high-resolution ZIMPOL imaging. This feature, located within 0.1$''$ of the star, represents the innermost spiral observed by high contrast imaging, a region typically obscured by the bright stellar halo. Figure \ref{fig:spiral_line} shows reconstructed flux distribution of the inner region around the star in narrow-band H$\alpha$, revealing an asymmetric spiral-like structure in both observations in 2016 and 2025. The flux reconstructed maps in H$\alpha$ are already presented in Sect. \ref{sec:results}, but here we show the scaled and continuum-subtracted images to show pure line H$\alpha$ features, then these images are convolved again with the PSFs to take into account the instrumental blurring.
As a side note, the images appear noisier, compared to single images, due to the subtraction step which combines noises in both input images.
It is noticeable that the spiral is not seen in continuum H$\alpha$, suggesting that this is strong line emissions. 
Dust was observed in this region, as traced by radio continuum as faint filamentary structures in the work of \citet{casassusFlowsGasProtoplanetary2013}, but because of the high depletion factor, it could be too weak to reach our minimum detectable contrast in continuum H$\alpha$.

\subsubsection{Variations in morphology and intensity}

After a period of nine years, the structures appear to have evolved, both in flux contrast and in position. We overlay some noticeable features that are visible in each epoch in Fig. \ref{fig:spiral_line}. In white color, arrows dashed lines show ones which appear consistent throughout the epochs, with the rotation of B during this period taken into account, including arm-like (dashed lines) and local intensity enhancement/discontinuity (arrows) structures. Besides, there seems to exist extended features that are visible in only one epoch, which are denoted in green dashed lines. Notably, the local signal enhancement marked by dotted green line is observed only in the 2025 epoch and is comparatively bright with respect to the surroundings. Also,
in the 2025 epoch, the flux of the spiral-like feature increased to be 2--3 times that of 2016, where it was roughly 20--30\% the flux of B. 
In addition, the convolved images in Fig. \ref{fig:spiral_line} are deprojected to $\rho$ as a function of $\theta$ to better inspect the pattern with respect to the position of B, as illustrated in in Fig. \ref{fig:spiral_deprojected}. Here the annotations point to relatively similar positions of features indicated in Fig. \ref{fig:spiral_line}. 
Though not quite explicit, between the two epochs, there are hints of a co-rotation with B, which has made a motion of $\sim$260$^\circ$ counter-clockwise. These signs provide evidence of B being linked to the observed spiral-like feature. However, we note that there are still some mismatches between the two observations. The structures are not entirely duplicated, which could lead to an explication that they could also have interaction with other objects apart from B, such as CC.
In any case, at separation of around 12 au from the star, the spiral-like feature could reside in the expected warp transition zone between the inner and outer disk, where the inclination is expected to vary continuously \citep{casassusAccretionKinematicsWarped2015}. Lacking detailed model of the warp geometry, we cannot deproject the images to determine the actual pattern speed and dynamical timescales.

As H$\alpha$ traces hot, ionized gas, we interpret this spiral as dust-depleted accretion streams at the shock front. Accretion shocks form when infalling gas reaches the inner disk boundary or circumstellar environment at supersonic velocities, generating radiative shocks that heat the gas to temperatures reaching an order of several thousands K, sufficient to excite strong H$\alpha$ recombination lines \citep{rafikovPROTOPLANETARYDISKHEATING2016, szulagyiEffectsPlanetaryTemperature2017, aoyamaTheoreticalModelHydrogen2018}. These accretion flows are also expected to vary in various timescales depending on the interaction between the star and the surrounding environment \citep[see, e.g., the review of][]{fischerAccretionVariabilityGuide2023}. We can cite the fascinating example of the eccentric binary system DQ Tau, where gaseous shock waves were reported to be simulated with a spiral shape and tangential discontinuities \citep{gomezdecastroAKScoEvidence2013}. Another interesting point is, while B's mass rate does not change significantly between the two epochs, the accretion streams in 2025 have approximately three times the flux of that in 2016. This effect can be explained by the orbital phase of B. The companion was approaching its periastron in 2019, capturing a part of the accretion flows, and in 2025 it was moving toward the apoastron, allowing more accretion to flow to inner parts. Orbitally modulated accretion in high-eccentricity binary systems caused by the angular momentum lost at periapstron has been predicted by hydrodynamic models \citep{guntherCircumbinaryDiskEvolution2002, deval-borroModellingCircumbinaryGas2011} as well as reported in observations of systems such as DQ Tau and AK Sco \citep{castroAccretionIntercycleVariations2020, alqubelatCoordinatedSpaceandGroundbased2026}.

\subsubsection{Companion-disk interactions}
\label{subsec:interactions}

Our time-variable H$\alpha$ observations provide the first direct kinematic evidence of shock-heated gas streamers in a circumbinary system. The spiral geometry and its evolution over nearly a decade suggest an organized inflow rather than random turbulence, consistent with hydrodynamic predictions for tidally forced accretion in eccentric binaries \citep{arzamasskiyDiskAccretionDriven2018}.
Streamers of gas from the outer disk to inner region of HD~142527 through non-Keplerian ``filaments'' in HCO+ emission was suggested by \citet{casassusFlowsGasProtoplanetary2013}. 
These streamers act as gap-crossing conduits that funnel mass from the outer disk into the inner accretion zone, maintaining inner disk stability despite the star's high accretion rate.
They also brought an interesting example of similar kinematics in GG Tau, where shocked gas in the inner disk was detected via molecular hydrogen gas \citep{beckCIRCUMBINARYGASACCRETION2012}. The presence of a binary companion to explain the accretion streamers as in GG Tau was also proposed, even though the mass ratios differ in the two systems. Later, \citet{priceCircumbinaryNotTransitional2018} prove that the dynamical interaction with HD~142527 B is indeed what generates the gas-crossing streamers.

\begin{figure}[ht] 
    \centering
    \subfloat{\includegraphics[width=0.8\hsize]{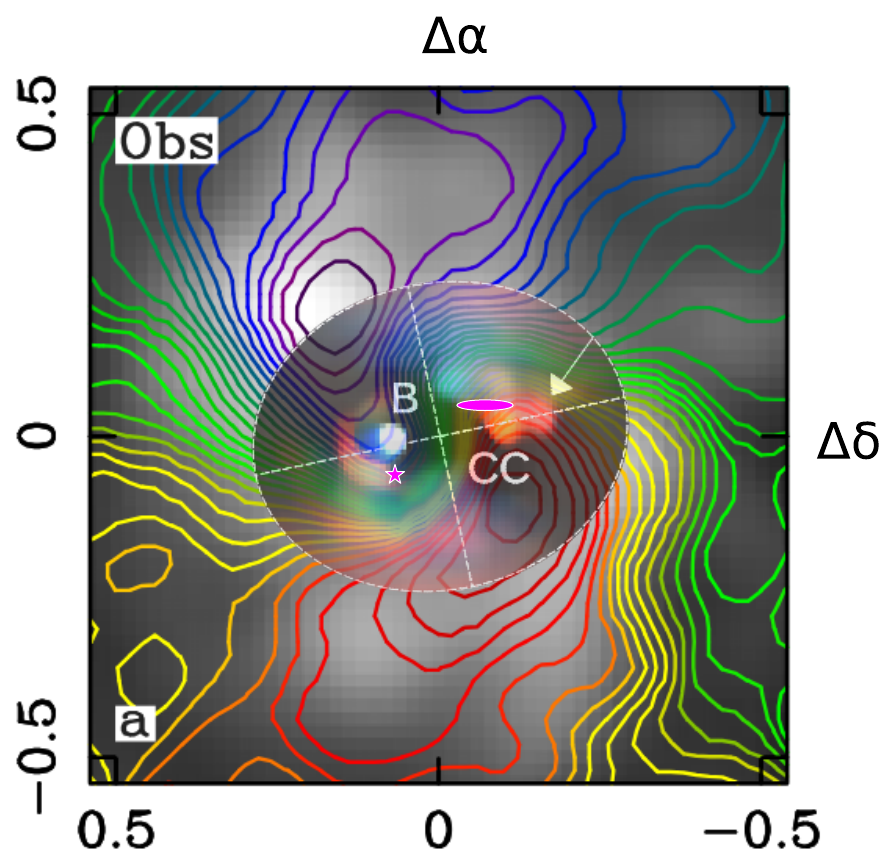}}
    \caption{\label{fig:warpdisk}ALMA CO(6–5) kinematics in the central region of HD~142527 \citep{casassusAccretionKinematicsWarped2015}, overlaid with reconstructed IFS flux from March 2016 in Fig. \ref{fig:reconstruction_irdifs}. The annotations show the predicted positions of B (magenta star symbol) and the CC (magenta ellipse) on 1 June 2012, when the ALMA observation was carried out.}
\end{figure}

We then explore the companion-disk interactions that could be inferred from our observations. It is interesting to look for connections among locations of the discovered spiral feature, the companions B, CC and the accretion kinematics of the system HD~142527. Non-Keplerian kinematics near the central star have been observed in the system, suggesting either fast radial flows or a warped disk going through disk tearing \citep{perezCOGASPROTOPLANETARY2014, rosenfeldFASTRADIALFLOWS2014, casassusAccretionKinematicsWarped2015, gargNonKeplerianSpiralsGaspressure2021}. These deviations from local Keplerian velocity pattern can most likely be explained by gravitational torques exerted by embedded planets or low-mass companions, with the strength depending on the mass of the perturbers \citep{rosenfeldFASTRADIALFLOWS2014, perezObservabilityPlanetdiscInteractions2018}. Figure \ref{fig:warpdisk} shows an overlay of the reconstructed IFS image in March 2016 (from Fig. \ref{fig:reconstruction_irdifs}) on the CO(6–5) velocity centroid map extracted from \citet{casassusAccretionKinematicsWarped2015}. Annotations indicate projected positions of B (magenta star symbol) and CC (magenta ellipse) from our orbit fit in Fig. \ref{fig:both_orbits} on June 1, 2012, the observing date of the reference centroid map. At the observing epoch, B and CC coincidentally lie at the nodes of the ``twisted'' velocity field. The gas seems to be pulled in two different directions simultaneously along the warp, with B and CC the at the center of the velocity disturbance. If CC is confirmed as a companion and the cause of the warped inner disk, alongside B, the system will become very intriguing, with multi-body interactions observed in both dust and gas across wavelength ranges. Then, the mechanism causing mutually inclined orbits between B and CC will be naturally clarified. Likewise, it will also explain the falling gas leading to shock heating and accretion linked with the spiral-like feature observed exactly at this active region in the vicinity of the star.

In summary, the ASDI-reconstructed flux distribution suggests that the spiral feature is tightly wound and closely linked to the gravitational influence of HD~142527 B and potentially also the newly detected CC. The accretion flows provide a probable explanation for the rapid mass infall required to sustain the high accretion rate of the primary star, that is through an accretion shock at the inner disk boundary, turning cold gas found in sub-millimeter observations to hot shocked gas at the shock front. The detected hot gas component, complement to the cold CO observations, unfolds the mechanism of energy dissipation when angular momentum turns to heat at the inner disk boundary. Continued H$\alpha$ flux monitoring and refined orbital constraints for CC will be essential to confirm the periodic variation of the spiral feature and clarify its relationship with both companions.

\section{Conclusions}
\label{sec:conclu}

Our study contains new insights into the inner region of the HD~142527 system. Using high-contrast imaging data obtained with the SPHERE instrument on VLT combined with advanced post-processing algorithms (\texttt{PACO} and \texttt{REXPACO} ASDI), we achieved unprecedented sensitivity to faint companions and disk structures within the system’s inner regions and made the following discoveries:
\begin{enumerate}
    \item{We revisited the low-mass companion HD~142527 B, and we provide updates on its photometry and astrometry. No significant accretion variability was detected. We also refined its orbit with the new astrometric measurements, verifying the high orbital eccentricity of the companion.}
    \item{We identified a CC to HD~142527 at $\sim$0.09$''$ in two epochs of NIR observations with IRDIS/IFS in 2016. Though we were not directly able to confirm it with an additional epoch in 2025, we acquired a tentative detection of a signal likely to be a CC. The CC’s red spectrum, combined with its close proximity to the star, suggests it may be embedded in the disk material, where it would contribute to the dynamical shaping of the inner disk. Nevertheless, the possibility of the CC being a disk feature cannot be ruled out.}
    \item {We reported the discovery of a spiral feature observed in two epochs in 2016 and 2025 with the narrow H$\alpha$ ZIMPOL filter. This accreting signal suggests hot gas flows inward to the inner disk around the star at the shock front. The finding of the spiral-like feature could explain the mass transfer mechanism of the system, and it complements the observations of cold gas in the literature. With hints of co-rotation with B, the feature provides a direct connection between companion dynamics and disk morphology. However, it is unclear if the feature interacts exclusively with B or also with other unseen bodies or structures, or even with the CC. }
\end{enumerate}

The HD~142527 system shares several characteristics with other well-studied protoplanetary disks, such as PDS 70 and AB Aur, where companion-driven disk features have been observed \citep[see, e.g.,][]{kepplerDiscoveryPlanetarymassCompanion2018, kepplerHighlyStructuredDisk2019, boccalettiPossibleEvidenceOngoing2020, currieImagesEmbeddedJovian2022, wahhajPDS70Unveiled2024}. However, unlike PDS 70, where the companions are of lower masses and located at larger separations (>20 au), B and the CC in the HD~142527 system are more massive and embedded within the inner disk, leading to more pronounced and complex interactions. The proximity and dynamical interaction of the CC and HD~142527 B offer a unique case study in understanding how multiple companions influence disk evolution.

The inner spiral structure possibly represents an accretion pathway channeling material from the outer disk toward the central star. This mechanism could contribute to the prolonged accretion phase observed in HD~142527, an essential factor in determining the final mass and orbital configuration of forming planets. The presence of multiple companions also raises the possibility of competitive accretion, where the gravitational influence of one companion affects the accretion rate and disk morphology associated with the other. Furthermore, the moderately high eccentricity of B and possibly also of CC places the star--companion--disk configuration in a high complexity, requiring extensive hydrodynamical models for better investigation of their interactions.

While this study addresses several longstanding questions about the HD~142527 system, it also raises new challenges and opportunities for further research. Ongoing and future observations will help figure out the true nature of the CC and confirm its orbital parameters. High-resolution imaging with VLT interferometry and observations with the James Webb Space Telescope or with next-generation instruments such as the Extremely Large Telescope could provide deeper insights into the disk's morphology and the physical properties of the CC. In addition, spectroscopic studies, particularly in the mid-infrared and submillimeter wavelengths, could reveal the chemical composition and temperature structure of the disk, offering clues about the ongoing planet formation processes. With these future observations, we will be able to clarify the accretion processes and the role of the companions in their interactions with the observed spiral structure. In addition, advanced modeling of the HD~142527 system, incorporating the newly detected CC and its interactions with the inner and outer disks, can provide a theoretical framework for interpreting the observed features.

\begin{acknowledgements}
We would like to thank the anonymous referee for the constructive report. 
We also thank H. M. Schmid for helpful discussions on the possible nature of the H$\alpha$ spiral-like feature and ZIMPOL data processing.
SPHERE is an instrument designed and built by a consortium consisting of IPAG (Grenoble, France), MPIA (Heidelberg, Germany), LAM (Marseille, France), LESIA (Paris, France), Laboratoire Lagrange (Nice, France), INAF Osservatorio di Padova (Italy), Observatoire de Genève (Switzerland), ETH Zurich (Switzerland), NOVA (Netherlands), ONERA (France) and ASTRON (Netherlands) in collaboration with ESO. 
SPHERE was funded by ESO, with additional contributions from CNRS (France), MPIA (Germany), INAF (Italy), FINES (Switzerland) and NOVA (Netherlands). 
SPHERE also received funding from the European Commission Sixth and Seventh Framework Programmes as part of the Optical Infrared Coordination Network for Astronomy (OPTICON) under grant number RII3-Ct-2004-001566 for FP6 (2004-2008), grant number 226604 for FP7 (2009-2012) and grant number 312430 for FP7 (2013-2016). This work has made use of the High Contrast Data Centre, jointly operated by OSUG/IPAG (Grenoble), PYTHEAS/LAM/CeSAM (Marseille), OCA/Lagrange (Nice), Observatoire de Paris/LESIA (Paris), and Observatoire de Lyon/CRAL, and supported by a grant from Labex OSUG@2020 (Investissements d’avenir – ANR10 LABX56). This work has been supported by the DDISK ANR contract number ANR-21-CE31-0015. This work was also supported by the Action Spécifique Haute Résolution Angulaire (ASHRA) of CNRS/INSU co-funded by CNES.
\end{acknowledgements}

\bibliographystyle{bibtex/aa} 
\bibliography{bibtex/bib} 

\begin{appendix}
\onecolumn

\section{Observations}
\label{sec:app_obs}

In this section we present the list of observations used in our analysis in Table \ref{tab:observations}.

\begin{table}[ht]
\centering
\caption{Summary of observations.}
\label{tab:observations}
\begin{tabular}{cccccccccc}
\hline \hline
\multirow{2}{*}{UT Date} & {Instr.} & {Filters\tablefootmark{a}} & {DIT $\times$ NDIT\tablefootmark{b}} & {N\textsubscript{exp}\tablefootmark{c}} & {Field rot.} & {Mean seeing}\tablefootmark{d} & {Strehl}\tablefootmark{d} & {$\tau_{0}$\tablefootmark{e}}\\
\multirow{2}{*}{~} & {~} & {~} & {(s)} & {~} & {(°)} & {($''$)} & {@1.6 $\mu$m} & {(ms)}\\
\hline

\multirow{4}{*}{2016-03-26} & IFS\tablefootmark{f} & YJH/ND1 & 4.0 $\times$ 56 & 16 & 65.24 & \multirow{2}{*}{0.69} & \multirow{2}{*}{0.70} & 1.37 \\
                            & IRDIS\tablefootmark{f} & K1--K2/ND1 & 0.84 $\times$ 88 & 32 & 64.86 &  & & 1.39 \\
                            & IFS & YJH/ND2 & 8.0 $\times$ 8 & 12 & 74.04 & \multirow{2}{*}{0.68} & \multirow{2}{*}{0.71} & 1.48 \\
                            & IRDIS & K1--K2/ND2 & 2.0 $\times$ 19 & 12 & 73.14 & & & 1.46 \\
\hline
\multirow{2}{*}{2016-03-30} & ZIMPOL & CntHa/N\_Ha & 30 $\times$ 10 & 7 & 48.66 & 0.28 & 0.79 & 2.73 \\
                            & ZIMPOL & CntHa/B\_Ha & 30 $\times$ 10 & 7 & 47.80 & 0.29 & 0.79 & 2.67 \\
\hline
\multirow{4}{*}{2016-06-13} & IFS\tablefootmark{f} & YJH/ND1 & 4.0 $\times$ 60 & 16 & 64.77 & \multirow{2}{*}{0.77} & \multirow{2}{*}{0.64} & 2.22 \\
                            & IRDIS\tablefootmark{f} & K1--K2/ND1 & 0.84 $\times$ 49 & 64 & 66.03 & & & 2.19 \\
                            & IFS & YJH/ND2 & 4.0 $\times$ 15 & 1 & ... & \multirow{2}{*}{0.84} & \multirow{2}{*}{0.60} & 1.90 \\
                            & IRDIS & K1--K2/ND2 & 2.0 $\times$ 1 & 1 & ... & & & 2.20 \\
\hline
\multirow{2}{*}{2025-04-27} & ZIMPOL & CntHa/N\_Ha & 30 $\times$ 15 & 7 & 64.11 & 0.27 & 0.78 & 4.71 \\
                            & ZIMPOL & CntHa/B\_Ha & 30 $\times$ 15 & 7 & 66.35 & 0.28 & 0.77 & 5.04 \\
\hline
\multirow{4}{*}{2025-05-09} & IFS\tablefootmark{f,g} & YJH/ND1 & 4.0 $\times$ 61 & 16 & 65.73 (63.56\tablefootmark{h}) & \multirow{2}{*}{0.38} & \multirow{2}{*}{0.86} & 15.32 \\
                            & IRDIS\tablefootmark{f,g} & K1--K2/ND1 & 0.84 $\times$ 50 & 64 & 67.28 & & & 15.32 \\
                            & IFS\tablefootmark{g} & YJH/ND2 & 8 $\times$ 8 & 2 & ... & \multirow{2}{*}{0.33} & \multirow{2}{*}{0.87} & ... \\
                            & IRDIS\tablefootmark{g} & K1--K2/ND2 & 2.0 $\times$ 19 & 2 & ... & & & 14.65 \\
\hline
\end{tabular}
\tablefoot{
\tablefoottext{a} {Filters and neutral density filters (if any) used for the observations.} 
\tablefoottext{b} {Detector integration time per frame $\times$ number of frames per exposure.} 
\tablefoottext{c} {Number of exposures.} 
\tablefoottext{d} {derived from data of ESO’s standard adaptive optics computing platform (\url{https://www.eso.org/sci/facilities/develop/ao/tecno/sparta.html}).}
\tablefoottext{e} {Coherence time.} 
\tablefoottext{f} {Saturated.}
\tablefoottext{g} {Low-wind effect.}
\tablefoottext{h} {After frame selection.} 
}
\end{table}
\FloatBarrier

\section{Supplementary observations}
\label{sec:app_supplement_obs}
In ESO archive of VLT/SPHERE system, we also acquired two more epochs of NIR observations of the system HD~142527 in 2015 (program ID 095.C-0298 -- PI Beuzit, J.-L). They are of the same observing configurations as the ones we used in Appendix \ref{sec:app_obs}, though with inferior conditions, field rotation and sensitivity. We decided to separated these observations from the better archival epochs (in 2016) that were presented as the main data of our work along with the new observations in 2025. Table \ref{tab:app_observations} shows information of these 2015 epochs.

\begin{table}[ht]
\centering
\caption{Summary of supplementary observations.}
\label{tab:app_observations}
\begin{tabular}{cccccccccc}
\hline \hline
{UT Date} & {Instr.} & {Filters} & {DIT $\times$ NDIT (s)} & {N\textsubscript{exp}} & {Field rot. (°)} & {Mean seeing ($''$)} & {Strehl @1.6 $\mu$m} & {$\tau_{0}$ (ms)}\\
\hline
\multirow{4}{*}{2015-05-05} & IFS\tablefootmark{a} & YJH/ND1 & 4.0 $\times$ 56 & 8 & 29.84 & 1.01 & \multirow{2}{*}{...} & 1.54 \\
                            & IRDIS\tablefootmark{a} & K1--K2/ND1 & 0.84 $\times$ 88 & 16 & 28.59 & 1.03 & & 1.52 \\
                            & IFS & YJH/ND2 & 8.0 $\times$ 8 & 8 & 40.52 & 1.09 & \multirow{2}{*}{...} & 1.45 \\
                            & IRDIS & K1--K2/ND2 & 2.0 $\times$ 19 & 8 & 39.47 & 1.05 &  & 1.41 \\
\hline
\multirow{4}{*}{2015-05-13} & IFS\tablefootmark{a} & YJH/ND1 & 4.0 $\times$ 7 & 8 & 25.95 & 0.60 & \multirow{4}{*}{0.8} & 8.99 \\
                            & IRDIS\tablefootmark{a} & K1--K2/ND1 & 0.84 $\times$ 88 & 16 & 30.35 & 0.61 & & 8.87 \\
                            & IFS & YJH/ND2 & 8.0 $\times$ 8 & 8 & 40.27 & 0.60 & ~ & 8.86 \\
                            & IRDIS & K1--K2/ND2 & 2.0 $\times$ 19 & 8 & 39.53 & 0.58 & & 9.25\\

\hline
\end{tabular}
\tablefoot{
\tablefoottext{a} {saturated}}
\end{table}

\begin{figure}[ht]
    \centering
    \includegraphics[width=\textwidth]{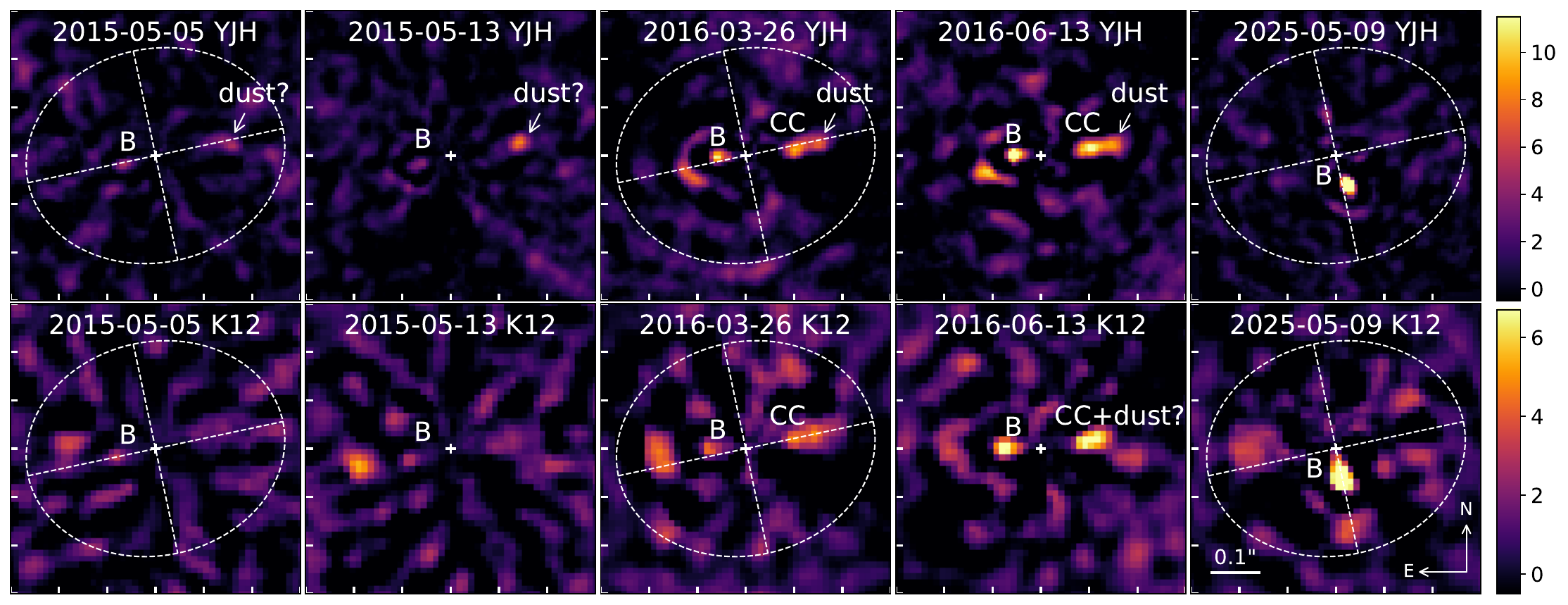}
    \caption{S/N detection maps of all IRDIS and IFS observations, combined with equal weights across spectral bands, obtained by \texttt{PACO} ASDI. The panels of 2016 and 2025 epochs are identical with which was shown in Fig. \ref{fig:snr_maps}. The dwarf companion HD~142527 B appears in all images, despite lower S/N in the first two epochs. In all the images, the star is marked with a white cross. An ellipse is overlaid in some panels for visual comparison purposes.}
    \label{fig:app_snr_irdifs}
\end{figure}

The processing of the supplementary data is identical to that of the main data in our work. Using \texttt{PACO}, we obtain the detection maps and put them together with that of later epochs to compare in Fig. \ref{fig:app_snr_irdifs}. We also detect B in all of the additional epochs with significant S/N (>5), though a little lower in $YJH$ filters (S/N 3.8) in the second epoch (Table \ref{tab:app_params}). There are also an enhancement of signal not far from the CC-dust combination detected position in 2016 epochs, which we suspect to be dusty material also (see also Sect. \ref{sec:results}). CC is not significantly detected in these two additional epochs. 

\begin{figure}[ht]
    \centering
    \subfloat{\includegraphics[width=0.5\textwidth]{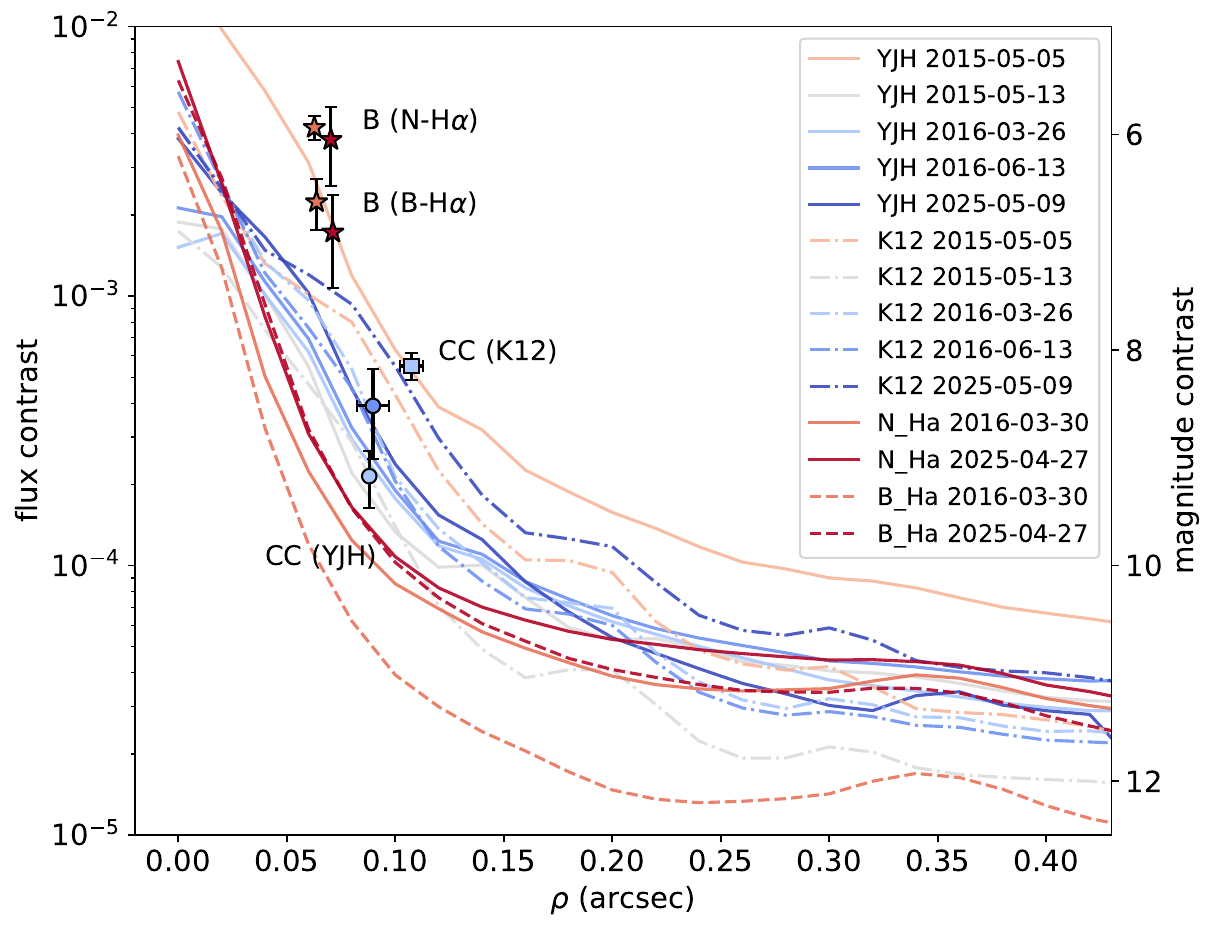}}
    \caption{\label{fig:app_contrast_curves}5$\sigma$ contrast curves of all observations mentioned in the study, including those in the main text and supplementary ones, overlaid by detected positions of B and CC in 3$\sigma$ in 2016 epochs.}
\end{figure}

In the same way as presented above, the contrast curves of all the observations are illustrated in Fig. \ref{fig:app_contrast_curves}. Not surprisingly, the sensitivity of the images in early May 2015 is not sufficient to be able to observe CC, while it could be the case in the second epoch in 2015. However, we note that, with significantly lower rotation field compared to that in 2016 epochs, it is unlikely that CC can be robustly detected with a high S/N value in these observations.

\begin{table}[ht]
\caption{\label{tab:app_params}Astrometric parameters of detected sources.}
\centering
\begin{tabular}{lccccccc}
\hline \hline
        \multirow{2}{*}{UT Date} & \multirow{2}{*}{Filters} & \multicolumn{3}{c}{B} &  \multicolumn{3}{c}{Dust} \\
        ~ & ~ & S/N & $\rho$ (mas) & $\theta$ (deg) & S/N & $\rho$ (mas) & $\theta$ (deg)  \\ \hline
        \multirow{2}{*}{05 May 2015} & $YJH$ & 5.1 & 82.7 ± 0.4 & 110.29 ± 0.24& 4.9 & 141.5 ± 0.6 & 275.92 ± 0.24 \\
        & $K12$ & 6.2 & 77.6 ± 0.2 & 107.56 ± 0.15&  ... &  ... & ... \\
        \multirow{2}{*}{13 May 2015} & $YJH$ & 3.8 & 73.8 ± 0.6 & 108.57 ± 0.43 & 8.2 & 141.3 ± 0.8 & 277.19 ± 0.30\\
        & $K12$ & 6.9 & 82.2 ± 0.2 & 108.88 ± 0.12&  ... &  ... & ...\\
        \hline
    \end{tabular}
\end{table}
\FloatBarrier

\section{Corner plot for the posterior distributions of HD~142527 B’ spectrum fit}
\begin{figure*}[ht] 
    \centering
    \includegraphics[page=2,width=1\hsize]{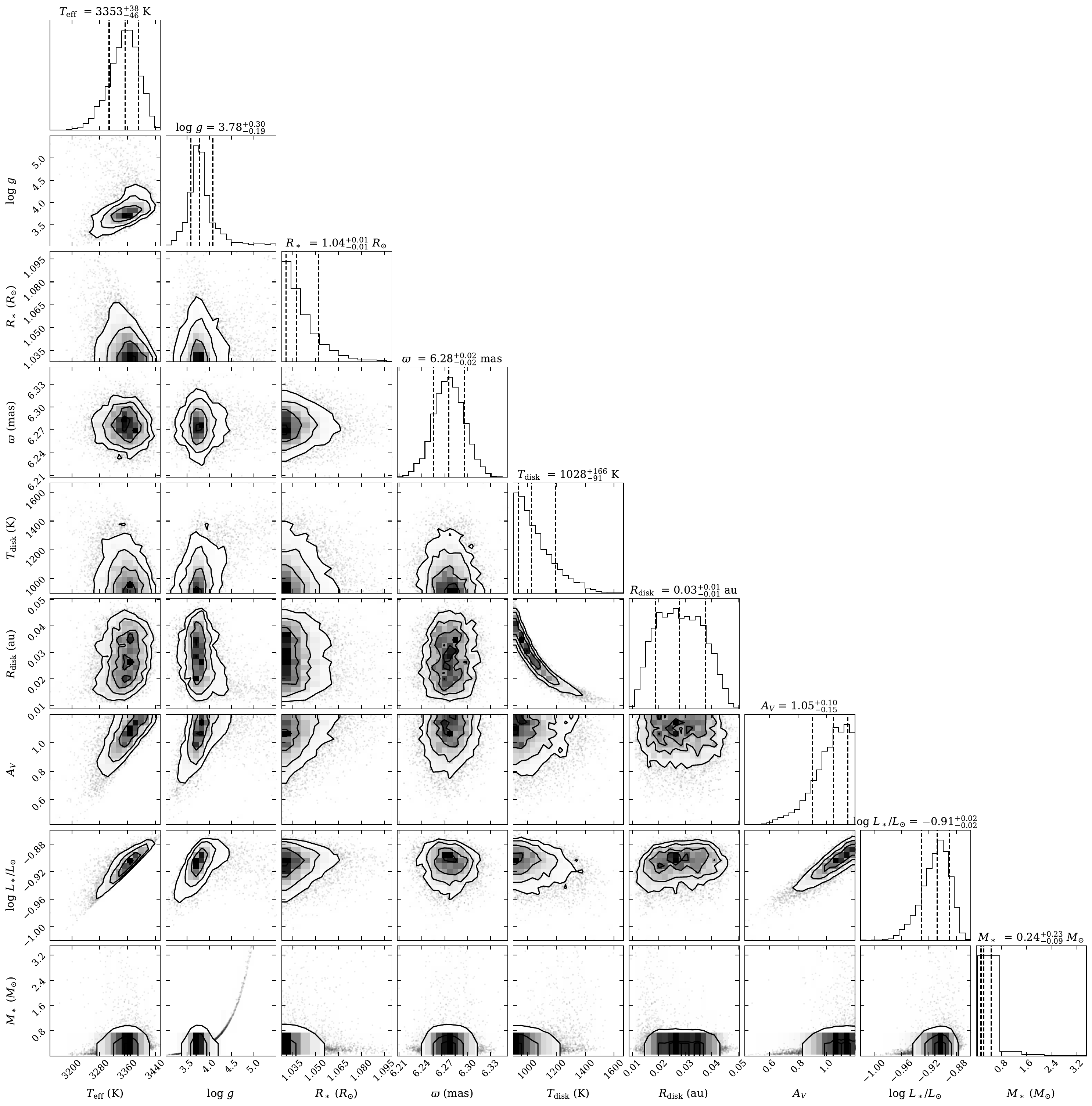}
    \caption{\label{fig:app_phot_B}Corner plot for the posterior distributions of HD~142527 B's spectrum fit, showing the 16\textsuperscript{th}, 50\textsuperscript{th} and 84\textsuperscript{th} percentiles for each of its parameters and the best-fit values obtained from median. 
    }
\end{figure*}

This section displays the corner plot for the posterior distributions of the HD~142527 spectrum fit in Fig. \ref{fig:app_phot_B}.
\label{sec:app_phot_B}
\FloatBarrier

\section{Corner plot for the posterior distributions of HD~142527 B’s orbit fit}
\begin{figure*}[ht] 
    \centering
    \includegraphics[width=1\hsize]{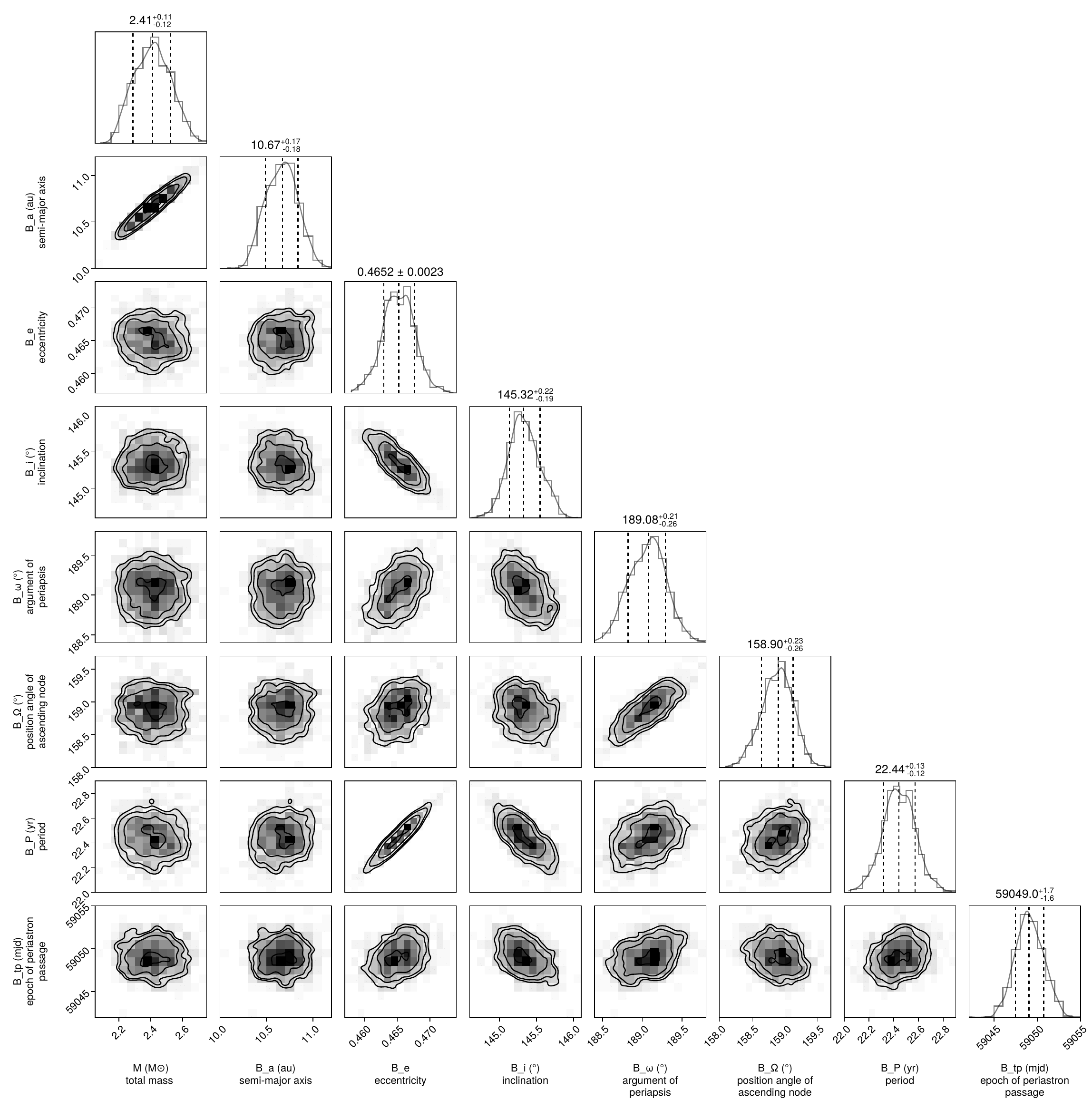}
    \caption{\label{fig:app_B_corner}Same as Fig. \ref{fig:app_phot_B} but for the HD~142527 B's orbit fit. Lacking of radial velocity data, there are supposed to be two peak values for each of B's $\omega$ and $\Omega$. However, we chose to display one peak for better visibility and comparison to parameters obtained from previous works.
    }
\end{figure*}

Figure \ref{fig:app_B_corner} illustrates the probability distributions of the orbit fitting parameters of HD~142527 B, assuming only one companion in the system
\label{sec:app_orbit_B}
\FloatBarrier

\section{Corner plot for the posterior distributions of HD~142527 B and CC’s simultaneous orbit fit}
\begin{figure*}[ht] 
    \centering
    \includegraphics[width=1\hsize]{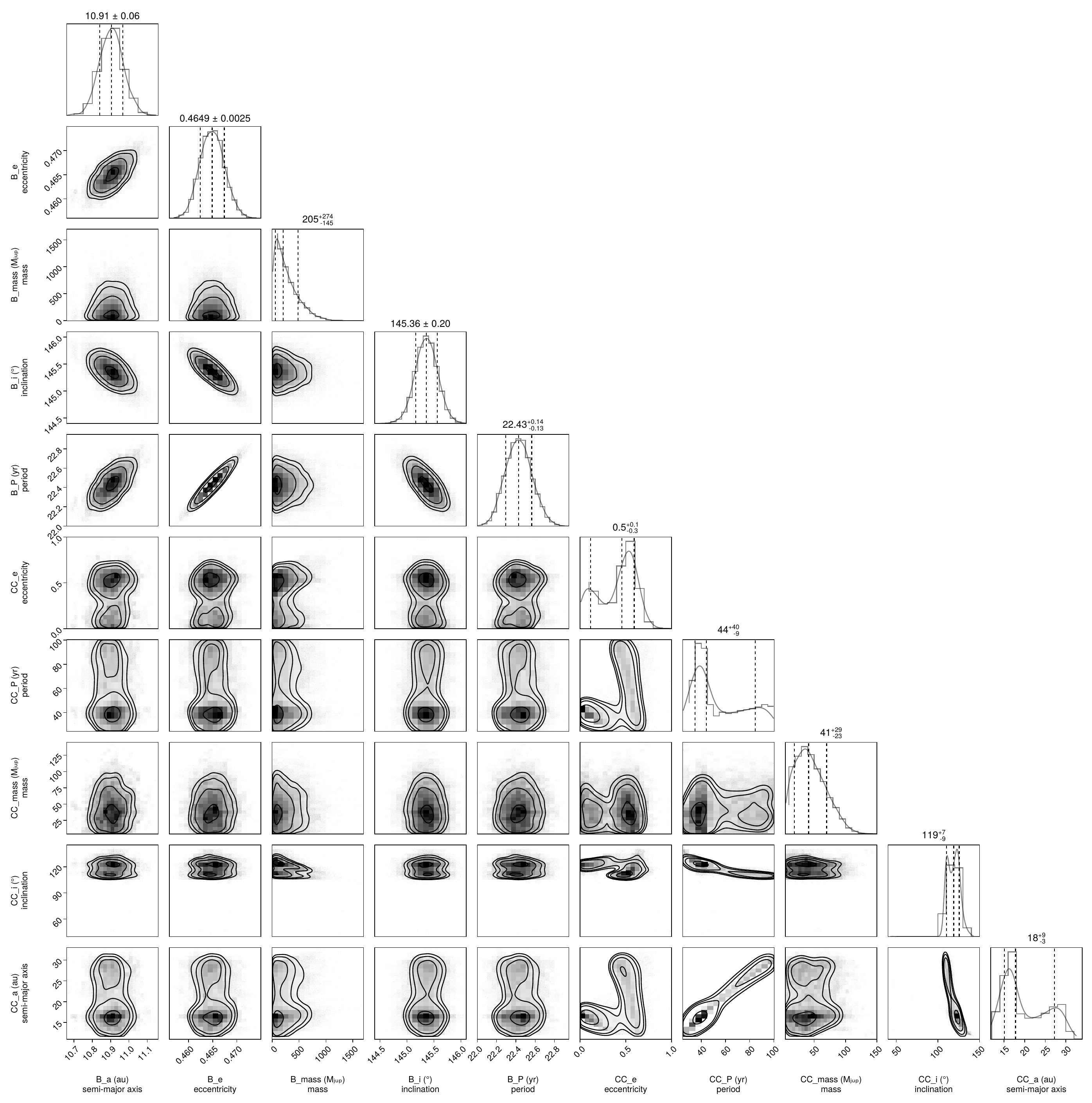}
    \caption{\label{fig:app_both_corner}Same as Fig. \ref{fig:app_phot_B} but for the orbit fit of both the CC and HD~142527 B simultaneously.
    }
\end{figure*}

 Figure \ref{fig:app_both_corner} illustrates the probability distributions of the orbit fitting parameters of both HD~142527 B and CC, where a third-epoch measurement is assumed to be the putative signal reported in 2025.
\label{sec:app_orbit_all}

\FloatBarrier

\end{appendix}

\end{document}